\documentclass[11pt,a4paper]{article}
\pdfoutput=1
\usepackage{jheppub}
\usepackage{tikz-feynman}

\usepackage{float}
\usepackage{amsmath}
\usepackage{cleveref}
\usepackage{xcolor}
\usepackage{caption}
\usepackage{subcaption}
\usepackage{physics}
\usepackage{tikz}
\usepackage{pgfplots}
\usepackage{multirow}

\crefname{equation}{eqn.}{eqns.}
\Crefname{equation}{Eqn.}{Eqns.}
\crefname{figure}{fig.}{figs.}
\Crefname{figure}{Fig.}{Figs.}

\newcommand{\be}{\begin{eqnarray}}
\newcommand{\ee}{\end{eqnarray}}

\usepackage{mathdots}
\usepackage{arydshln}

\title{Four results on out-of-equilibrium 2PI simulations in 3+1 dimensions}

\author[a]{Anders Tranberg,}
\author[a]{Gerhard Ungersb\"ack,}

\affiliation[a]{Faculty of Science and Technology, University of Stavanger, 4036 Stavanger, Norway}

\emailAdd{anders.tranberg@uis.no}
\emailAdd{gerhard.ungersback@uis.no}

\abstract{We perform an analysis of a number of approximations and methods used in numerical  simulations of  real-time Kadanoff-Baym equations  based on truncations of the 2PI effective action. We compare the loop expansion to the 1/N expansion and compare their classical limit to classical-statistical simulations. We also compare implementations based on a space-time lattice discretization at the level of the action to an ad hoc momentum discretization at the level of the equations of motions. We extract some rules of thumb for performing 2PI-simulations of out-of-equilibrium systems.}

\begin{document}

\maketitle

\section{Introduction}
\label{sec:Intro}

Out-of-equilibrium quantum evolution is at the core of many phenomena in early Universe cosmology, in the early stages of heavy ion collisions and many condensed matter systems. The real-time sign problem makes direct evaluation of observables through the path integral extremely challenging, although substantial effort has been invested into developing alternative ways of performing importance sampling, most notably the complex Langevin method \cite{SQ1,SQ2,SQ3,SQ4,SQ5} and Picard-Lefschetz thimbles \cite{PL1,PL2,PL3,PL4}. The current state of the art simulations are limited to quantum mechanical systems and very short time extents \cite{SQ6}, or field theory at small coupling \cite{SQ7}. Although these attempts are promising, at present real-world applications remain out of reach. In order to gain insight into out-of-equilibrium processes one must therefore introduce approximations. 

Two well established approaches are the classical-statistical approximation to out-of-equilibrium field theory (following for instance \cite{Aarts:1997kp,Millington}) and Schwinger-Dyson/Kadanoff-Baym evolution of correlators based on truncations of the 2PI effective action \cite{Cornwall:1974vz,BergesCox,Aarts:2001qa}. The former does not rely on any expansion or truncation, and is therefore accurate to all orders in the coupling. Since it is the classical limit, all truly quantum effects are however absent, and it will only give accurate results when such effects are subdominant. Examples of where the approximation fails include incorrect thermalisation and late-time state \cite{Aarts:2001yn,Arrizabalaga:2004iw} and the absence of instantons and quantum tunneling \cite{Hertzberg:2020tqa,Tranberg:2022noe}.

On the other hand, the 2PI formalism includes both quantum and classical effects, for instance producing the correct late-time thermal state, but relies on a truncation of diagrammatic expansions of the effective action. Results are therefore accurate only up to some order in an expansion parameter in terms of precision, but also that some qualitative effects are absent at a given order. Examples include the absence of thermalisation of scalar fields at LO in a coupling or 1/N expansion, and the absence of topological defects for scalar fields at NLO \cite{Rajantie:2006gy,Berges:2010nk}. 

For scalar field theory with O(N) symmetry, on which we focus in this work, two diagrammatic expansions exist: The loop expansion \cite{BergesCox} and the $1/N$ expansion \cite{Aarts:2002dj,Berges:2001fi,Aarts:2008wz,Aarts:2006cv}. Both have been studied and applied extensively, see \cite{App0,App1,App2,App3,App4,App5,App6,App7,App8,App9,App10,App11,App12,App13,App14,App15,App15a,App16,App16a,App17,App18,App19,App20,App21,App22,App23,App24,App25,App26}.

In the following we analyse both expansions truncated at NLO, since this is the first order that allows for non-trivial evolution (damping, dissipation, thermalisation). As our measurement of choice we consider the relaxation of a displaced momentum mode away from the (near-)thermal background initial state. This provides a damping rate, which we are able to compare across approximations and parameter values. 

Conveniently, the 2PI approach has a simple classical limit, for which we have the exact result for comparison, the classical-statistical approximation. We will make use of this to assess the accuracy of the two expansions and truncations.

 In the literature one also finds two different discretisation methods for out-of-equilibrium 2PI simulations. The first, derived from a lattice action in which the space-time is discretised on a $N_x^3 \times N_t$ hypercubic lattice and the second where the discretisation is imposed at the level of the equations of motion. We will compare the two, and identify the resolution (lattice spacing, UV cut-off) at which they give matching results.

The paper is organized as follows: In section \ref{sec:Model} we introduce the 2PI formalism. Subsections are devoted to the different expansions and the numerical implementation. In section \ref{sec:CompLaN} we introduce how we quantify equilibration properties and compare the aforementioned expansions. In section \ref{sec:ClassComp} we discuss the classical limit, and how different expansions compare. Section \ref{sec:ClasTherm} focuses on the accuracy of simulations in terms of the final states. This is again done in the classical limit and the difference to the full quantum simulation is demonstrated and discussed. In section \ref{sec:Discret} we compare different discretisation schemes and finally draw some conclusions in section \ref{sec:Conclusion}.

\section{Model and evolution equations}
\label{sec:Model}
We will consider a system of $N$ real, self-interacting scalar fields with continuum action in 3+1 space-time dimensions
\begin{eqnarray}
S=\int d^4x \left[
\frac{1}{2}\partial_\mu\phi_a\partial^\mu\phi_a-\frac{m^2}{2}\phi_a\phi_a-\frac{\lambda}{24 N}(\phi_a\phi_a)^2
\right],\qquad a=1,\ldots, N.
\end{eqnarray}
We will assume that $\lambda >0$ and that $m^2>0$ so that there is no symmetry breaking. Time is defined along the Schwinger-Keldysh contour. We will briefly describe how the real-time Kadanoff-Baym equations for the propagator arises, based the 2PI effective action, which we will truncate at NLO in both a 2PI loop expansion and a 1/N-expansion. More details may be found in the literature \cite{BergesCox,Aarts:2002dj}.
The 2PI effective action may be written in the form \cite{Cornwall:1974vz}:
\begin{equation}
    \Gamma[\bar{\phi},G]=S[\bar{\phi}]-\frac{i}{2} \rm Tr \ln{G}+\frac{i}{2} \rm Tr \, G_{0}^{-1}[\bar{\phi}]G+\Phi[\bar{\phi},G].
\end{equation}
It is a functional of the mean field and the propagator, defined by
\begin{eqnarray}
\bar{\phi}_a(t)=\langle\phi_a({\bf x},t)\rangle,\qquad G_{ab}({\bf x-y},t,t')= \langle T_\mathcal{C}\phi_a({\bf x},t)\phi_b({\bf y},t)\rangle,
\end{eqnarray}
where $T_\mathcal{C}$ denotes time-ordering along the Schwinger-Keldysh contour, 
and involves the classical inverse propagator
\begin{align}
    iG^{-1}_{0,\,ab}(x,y)=
    \frac{\delta^{2}\,S[\bar{\phi}]}{\delta\bar{\phi}_{a}(x)\,\bar{\phi}_{b}(y)}.
\end{align}
The object $\Phi[\bar{\phi},G]$ denotes the rest of the effective action and may be written as an expansion in terms of skeleton Feynman diagrams, but with fully dressed progagators $G$ and mean fields $\bar{\phi}$, ultimately to be solved for self-consistently. 

Variation of the 2PI effective action with respect to the propagator and mean field give their real-time equations of motion. For simplicity, we will consider a homogeneous initial state, where we also set the mean field to zero. It follows that the mean field remains zero throughout the evolution and therefore we are not concerned with its equation of motion. For the propagator we find
\begin{align}
    \frac{\delta \Gamma[\bar{\phi},G]}{\delta G_{ab}(x,y)}=0
    \to
    G^{-1}_{ab}(x,y)=G_{0,ab}^{-1}(x,y)+i\Sigma_{ab}(x,y),
    \label{eq:G1}
\end{align}
where we introduced the self-energy
\begin{align}
    \Sigma_{ab}(x,y) =
    -2\frac{\delta \Phi[\bar{\phi},G]}{\delta G_{ab}(x,y)}.
    \label{eq:SelfE}
\end{align}
The propagator is decomposed in terms of the statistical and spectral components via
\begin{align}
G_{ab}(x,y)= F_{ab}(x,y) - \frac{i}{2} \rho_{ab}(x,y) \textrm{sign}_{\mathcal{C}}(x^0-y^0),
\label{eq:Gdecomp}
\end{align}
defined through 
\begin{align}
F(x,y)\delta_{ab}=F_{ab}(x,y) = \frac{1}{2} \langle \{\phi_a(x), \phi_b(y)\} \rangle
,\qquad \rho(x,y)\delta_{ab}=\rho_{ab}(x,y) = i \langle [\phi_a(x), \phi_b(y)] \rangle,
\end{align}
where the $O(N)$ symmetry implies that the propagators are diagonal in colour-space $\propto \delta_{ab}$.
The equation of motion then reduces to explicit evolution equations for $F$ and $\rho$, 
\begin{align}
        \Big( \partial_t^2 - \partial_x^2 + M^2(x) \Big) F(x,y) =
        &- \int_0^{x^0} dz^0 \int d^3z \,\Sigma^{\rho}(x,z) F(z,y) \nonumber \\ 
        &+ \int_0^{y^0} dz^0 \int d^3z \,\Sigma^{F}(x,z) \rho(z,y),\\
        \Big( \partial_t^2 - \partial_x^2 + M^2(x) \Big) \rho(x,y) =
        &- \int_{y^0}^{x^0} dz^0 \int d^3z \,\Sigma^{\rho}(x,z) \rho(z,y).
\label{eq:eom}
\end{align}
Any local (in time) contributions to the self-energy are gathered into a time dependent mass $M(x)$, while the remaining non-local (in time) contributions enter in the memory integrals on the right-hand side, stretching all the way back to the initial time. 

Finally, we need to specify a diagrammatic expansion and a truncation, to provide a self-energy as specified in (\ref{eq:SelfE}). In the following we will consider the loop and $1/N$ expansions at NLO.

\subsection{Loop expansion}
\label{sec:lambdaexp}

The diagrammatic expansion in loop order, or equivalently in superficial\footnote{Since the full self-consistent propagators include resummation to all orders in $\lambda$.} orders of $\lambda$, to NLO is given by
\begin{align}
&\Phi_{\lambda}[G] = 
    - \frac{\lambda}{24 N} \int d^{4}x\, \left[G_{aa}(x,x)G_{bb}(x,x)+2G_{ab}(x,x)^2 \right]\nonumber\\
    &+ i \left(\frac{\lambda}{12N}\right)^2 \int d^{4}x \int d^{4}y\, \left[G_{ab}(x,y)^2G_{cd}(x,y)^2+2G_{ab}(x,y)G_{bc}(x,y)G_{cd}(x,y)G_{da}(x,y)\right], 
\label{eq:Philambda}
\end{align}
where the first term is LO $\mathcal{O}(\lambda)$ and local in time, and the second term is NLO $\mathcal{O}(\lambda^2)$ and non-local in time. The diagrams are shown in Figure \ref{fig:lambdadiagrams}. The expressions for the self-energy and mass become:
\begin{align}
M^2(x) &= m^2 + \lambda \frac{N+2}{6N} F(x,x),  \\
\Sigma^{\rho}(x,y) &= - \lambda^2 \frac{N+2}{18 N^2}  \rho(x,y) \Big( 3F(x,y)^2 - \frac{1}{4} \rho(x,y)^2 \Big), \\
\Sigma^{F}(x,y)    &= - \lambda^2 \frac{N+2}{18 N^2} F(x,y) \Big( F(x,y)^2 - \frac{3}{4} \rho(x,y)^2 \Big).
\label{eq:NLOlambda}
\end{align}
We see that LO in a coupling expansion involves contributions at $\mathcal{O}(1)$ and $\mathcal{O}(1/N)$ in the self-energy, while at NLO we have contributions at $\mathcal{O}(1/N)$ and $\mathcal{O}(1/N^2)$. The expansion is equally valid for any value of $N$.

\subsection{$1/N$ expansion}
\label{sec:Nexp}

The $1/N$ expansion at NLO is given by the following functional
\begin{align}
\Phi_{1/N}[G] &= - \frac{\lambda}{24 N} \int d^{4}x\, G_{aa}(x,x) G_{bb}(x,x) 
                + \frac{i}{2} \int d^{4}x\, \ln \textrm{B}(G) (x,x),
\label{eq:PhiNinverse}
\end{align}
where 
\begin{align}
    \textrm{B}(x,y; G) = \delta_{\mathcal{C}}(x-y) + i \frac{\lambda}{6N} G_{ab}(x,y) G_{ab}(x,y).
\end{align}
The diagrams are shown in Figure \ref{fig:Ninversediagrams}.
The expressions for the self-energy and mass are given by:
\begin{align}
    M^2(x) &= m^2 + \lambda \frac{N+2}{6 N} F(x,x), \\
    \Sigma^{\rho}(x,y) &= - \frac{\lambda}{3 N} \Big( F(x,y) I_{\rho}(x,y)     + \rho(x,y) I_{F}(x,y) \Big), \\
    \Sigma^{F}(x,y)    &= - \frac{\lambda}{3 N} \Big( F(x,y) I_{F}(x,y) - \frac{1}{4} \rho(x,y) I_{\rho}(x,y) \Big),
    \label{eq:NLON}
\end{align}
where now the $I_{F,\rho}$ themselves involve memory integrals
\begin{align}
    I_{\rho}(x,y) &=  \frac{\lambda}{3} \Big( F(x,y) \rho(x,y) \Big)
    - \frac{\lambda}{3} \int_{y^0}^{x^0} dz\, I_{\rho}(x,z) \big( F(z,y)\rho(z,y) \big), \\
    I_F(x,y) &=  \frac{\lambda}{6} \Big( F(x,y)^2 -\frac{1}{4} \rho(x,y)^2 \Big)
    - \frac{\lambda}{6} \Big( 
    \int_0^{x^0} dz\, I_{\rho}(x,z) \big( F(z,y)^2 -\frac{1}{4} \rho(z,y)^2 \big) \\
    &- 2 \int_0^{y^0} dz\, I_{F}(x,z) F(z,y) \rho(z,y) \Big).
    \label{eq:NLONI}
\end{align}
We note that the second LO term of the $\lambda$ expansion now appears at NLO in the $1/N$ expansion. Similarly, diagrams at all orders in $\lambda$, but at the same order in $1/N$, are resummed by the $I_{F,\rho}$ auxiliary propagators.
\begin{figure}
   \centering
\begin{subfigure}[b]{0.2\textwidth}
    \begin{equation*}
        \Phi_{\lambda}^{\rm LO}[G] =
        \begin{gathered}
        \begin{tikzpicture}[line width=.8pt]
        \begin{feynman}
                \vertex (v);
                \vertex[above=1.0cm of v](t);
                \vertex[below=1.0cm of v](b);
                \diagram*{
                (v)  -- [out=180,in=180, looseness=1.75] (t) -- [ out=0,in=0, looseness=1.75] (v)
                     -- [out=-180,in=180, looseness=1.75] (b) --[ out=0,in=0, looseness=1.75] (v)
                };
                \draw[fill=black] (v) circle (1pt);
        \end{feynman}
        \end{tikzpicture}
        \end{gathered}
    \end{equation*}
\end{subfigure}
\begin{subfigure}[b]{0.3\textwidth}
    \begin{equation*}
        , \quad \Phi_{\lambda}^{\rm NLO}[G] =
        \begin{gathered}
        \begin{tikzpicture}[line width=.8pt]
        \begin{feynman}
            \vertex (v1);
            \vertex[right=1.5cm of v1] (v3);
            \diagram* {
                (v1) -- [ half left, looseness=1.7] (v3) -- [  half left, looseness=1.7] (v1)
                };
            \diagram* {
                (v1) -- [ half left, looseness=0.7] (v3) -- [  half left, looseness=0.7] (v1)
                };
            \draw[fill=black] (v1) circle (1pt);
            \draw[fill=black] (v3) circle (1pt);
        \end{feynman}
        \end{tikzpicture}
        \end{gathered}
    \end{equation*}
\end{subfigure}
    \caption{Feynman diagrams corresponding to LO (left) and NLO (right) in an $\lambda$ expansion. The diagrams are built from self-consistently dressed propagators, resumming an infinite set of perturbative diagrams.}
    \label{fig:lambdadiagrams}
\end{figure}

\begin{figure}
\begin{subfigure}[b]{0.2\textwidth}
    \begin{equation*}
        \Phi_{\rm 1/N}^{\rm LO}[G] =
        \begin{gathered}
        \begin{tikzpicture}[line width=.8pt]
        \begin{feynman}
                \vertex (v);
                \vertex[above=1.0cm of v](t);
                \vertex[below=1.0cm of v](b);
                \diagram*{
                (v)  -- [out=180,in=180, looseness=1.75] (t) -- [ out=0,in=0, looseness=1.75] (v)
                     -- [out=-180,in=180, looseness=1.75] (b) --[ out=0,in=0, looseness=1.75] (v)
                };
                \draw[fill=black] (v) circle (1pt);
        \end{feynman}
        \end{tikzpicture}
        \end{gathered}
    \end{equation*}
\end{subfigure}
\begin{subfigure}[b]{0.3\textwidth}
    \begin{equation*}
        , \quad \Phi_{\rm 1/N}^{\rm NLO}[G] =
        \begin{gathered}
        \begin{tikzpicture}[line width=.8pt]
        \begin{feynman}
                \vertex (v);
                \vertex[above=1.0cm of v](t);
                \vertex[below=1.0cm of v](b);
                \diagram*{
                (v)  -- [ out=180,in=180, looseness=1.75] (t) --[ out=0,in=0, looseness=1.75] (v)
                     -- [ out=-180,in=180, looseness=1.75] (b) --[  out=0,in=0, looseness=1.75] (v)
                };
                \draw[fill=black] (v) circle (1pt);
        \end{feynman}
        \end{tikzpicture}
        \end{gathered}
    +        
        \begin{gathered}
        \begin{tikzpicture}[line width=.8pt]
        \begin{feynman}
            \vertex (v1);
            \vertex[right=1.5cm of v1] (v3);
            \diagram* {
                (v1) -- [ half left, looseness=1.7] (v3) -- [  half left, looseness=1.7] (v1)
                };
            \diagram* {
                (v1) -- [ half left, looseness=0.7] (v3) -- [  half left, looseness=0.7] (v1)
                };
            \draw[fill=black] (v1) circle (1pt);
            \draw[fill=black] (v3) circle (1pt);
        \end{feynman}
        \end{tikzpicture}
        \end{gathered}
    +
        \begin{gathered}
        \begin{tikzpicture}[line width=.8pt]
        \begin{feynman}
            \vertex (v1);
            \vertex[right=1.5cm of v1] (v3);
            \vertex[above right=1.06cm of v1] (v2); 
            \diagram* { 
                (v1) -- [ half left, looseness=1.7] (v3) -- [  half left, looseness=1.7] (v1)
                };
            \diagram* { 
                (v1) -- [ half right, looseness=0.5] (v2) -- [ half right, looseness=0.5] (v3) -- [ half left, looseness=0.7] (v1)
                };
            \draw[fill=black] (v1) circle (1pt);
            \draw[fill=black] (v2) circle (1pt);
            \draw[fill=black] (v3) circle (1pt);
        \end{feynman}
        \end{tikzpicture}
        \end{gathered}
    +
        \begin{gathered}
        \begin{tikzpicture}[line width=.8pt]
        \begin{feynman}
            \vertex (v1);
            \vertex[right=1.5cm of v1] (v3);
            \vertex[above right=1.06cm of v1] (v2); 
            \vertex[below right=1.06cm of v1] (v4); 
            \diagram* { 
                (v1) -- [ half left, looseness=1.7] (v3) -- [ half left, looseness=1.7] (v1)
                };
            \diagram* { 
                (v1) -- [ half right, looseness=0.5] (v2) -- [ half right, looseness=0.5] (v3) -- [ half right, looseness=0.5] (v4) -- [ half right, looseness=0.5] (v1)
                };
            \draw[fill=black] (v1) circle (1pt);
            \draw[fill=black] (v3) circle (1pt);
            \draw[fill=black] (v2) circle (1pt);
            \draw[fill=black] (v4) circle (1pt);
        \end{feynman}
        \end{tikzpicture}
        \end{gathered}
    + \cdots
    \end{equation*}
\end{subfigure}
    \caption{Feynman diagrams corresponding to LO (left) and NLO (right) in an 1/N expansion. The diagrams are again built from self-consistently dressed propagators, resumming an infinite set of perturbative diagrams.}
    \label{fig:Ninversediagrams}
\end{figure}

\subsection{Initial conditions and observables}
\label{sec:IniConds}

The evolution equations are re-derived from an action defined on a space-time lattice with a benchmark size of $N_x^3=16^3$ spatial sites and with a benchmark lattice spacing of $am=0.7$. The time direction is discretized with a time-step of $dtm=0.07$, and we keep the memory $n_t=1000$ time-steps back, corresponding to a physical time extent of $n_t dt m=mt_{K}=70$.

We define the Gaussian \footnote{See \cite{App12} for generalisation to non-Gaussian initial conditions.}initial conditions in momentum space\footnote{Defining $\rho(x,y)=\int \frac{d^3k}{(2\pi)^3}\rho_k(t,t')e^{i{\bf k}({\bf x-y})}$, $F(x,y)=\int \frac{d^3k}{(2\pi)^3}F_k(t,t')e^{i{\bf k}({\bf x-y})}$}, where the spectral propagator at the initial time follows from the commutation relations
\begin{align}
        \rho_k(t,t) = 0, \quad \partial_t \rho_k(t,t^{\prime})|_{t=t^{\prime}} = 1.
\end{align}
The statistical propagator may for a free theory be written as
\begin{align}
        F_k(t,t^{\prime}) |_{t=t^{\prime}=0} &= \frac{1}{\omega_k} \Big(n_k + \frac{1}{2}\Big), \\ 
        \partial_t F_k(t,t^{\prime}) |_{t=t^{\prime}=0} &= 0 , \\
        \partial_t \partial_{t^{\prime}} F_k(t,t^{\prime}) |_{t=t^{\prime}=0} &= \omega_k \Big(n_k + \frac{1}{2}\Big). 
\end{align}
and we will define the initial conditions through a choice of $\omega_k$ and $n_k$.
\begin{figure}[ht]
  \centering
  \includegraphics[width=0.5\textwidth]{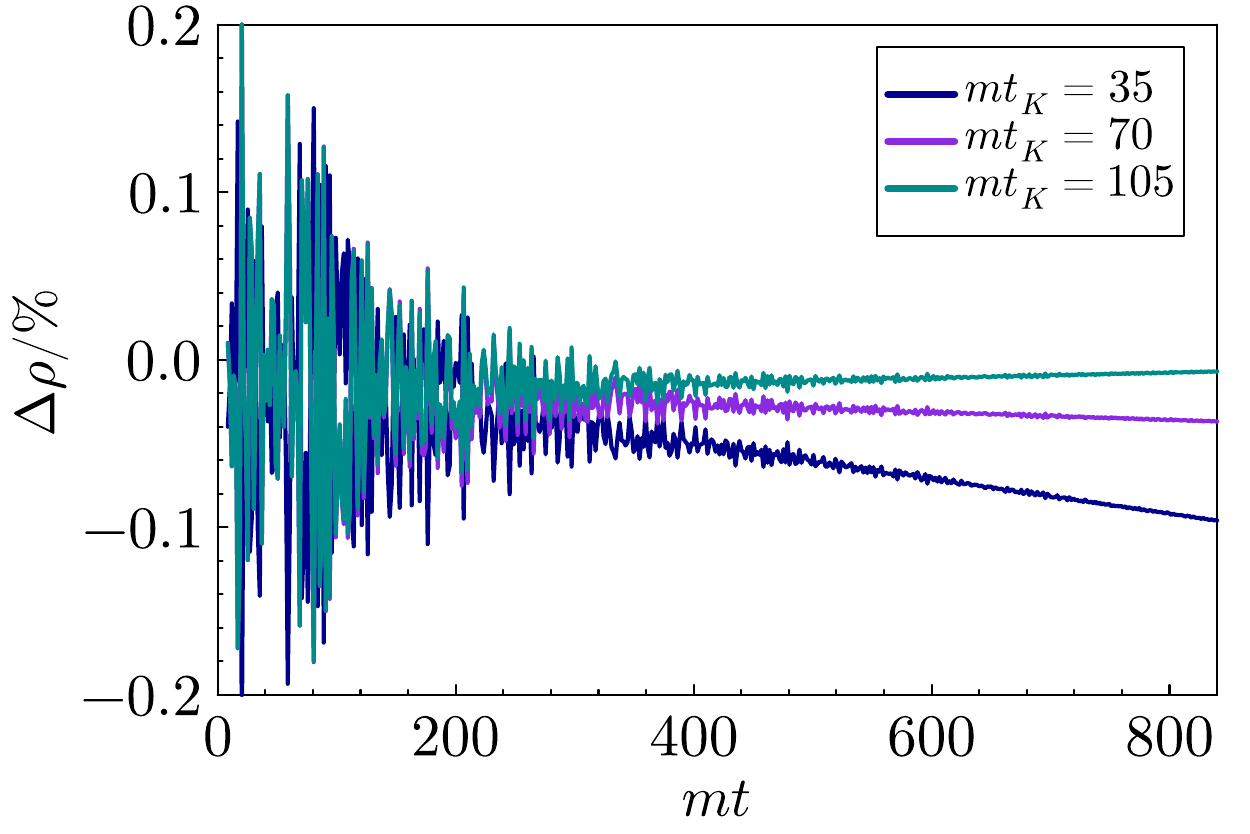}
  \caption{Energy loss due to cutoff of memory kernel at given time $mt_k$. We use $\lambda=10$, $N=10$ and $T/m=3$.}
  \label{fig:Eloss}
\end{figure}
Conversely, as long as the interactions are small enough that the fluctuations are quasi-particle like, we may extract the particle number $n_k$ and the dispersion relation\footnote{On the lattice, we further introduce a correction $n_k(t) + \frac{1}{2} \rightarrow c_k(t)(n_k(t) + \frac{1}{2})$, where \mbox{ $c_k(t) = \sqrt{ 1 - \frac{1}{4} dt^2 \omega_k^2(t) }$}, to account for time-discretization effects.} by inverting these relations:
\begin{align}
    \omega_k(t) &= \sqrt{ \frac{\partial_t \partial_{t^{\prime}} F_k(t,t^{\prime})|_{t^{\prime}=t}}{F_k(t,t)} } 
   \label{eq:disprelation},  \\
    n_k(t) + \frac{1}{2} &=  \sqrt{ \partial_t \partial_{t^{\prime}} F_k(t,t^{\prime})|_{t^{\prime}=t} F_k(t,t)}.
   \label{eq:particlenrdefinition},  \\
\end{align}
The evolution equations have the property that, up to oscillations due to lattice discretization, they conserve an energy functional 
\begin{align}
\label{eq:Eden}
    \frac{\rho(t)}{N} = 
    & \frac{1}{2} \int d^3 k \partial_t \partial_{t^{\prime}} F_k(t,t) 
     +\frac{1}{2} \int d^3 k  \Big( k^2 + m^2 + \frac{\lambda}{2} \frac{N+2}{6N} \int \frac{d^3k^{\prime}}{(2\pi)^3} F_{k^{\prime}}(t,t) \Big) F_k(t,t) \nonumber\\
    & +\frac{1}{4} \int_0^t dt^{\prime} \int d^3k \Big(\Sigma^{\rho}_k(t,t^{\prime}) F_k(t^{\prime},t) - \Sigma^{F}_k(t,t^{\prime}) \rho_k(t^{\prime},t) \Big).
\end{align}
The evolution equations drive the system towards thermal equlibrium, in the sense that for late times, any initial condition will equilibrate to a universal stationary state, defined only by the energy density. For small coupling it is a Bose-Einstein distribution with a thermally corrected effective mass.
The conservation of energy is accurate, provided the memory integrals stretch all the way back to the initial time.
However in practice, one must cut off the kernel at some finite time, which leads to a small net drift of the total energy. Figure \ref{fig:Eloss} shows the energy loss due to the reduced memory kernel. After an initial phase of fluctuations the value of the energy density stabilizes and we see that the longer the memory kernel is the less energy is drained from the system. We use  $mt_k=70$ throughout in this article to obtain robust damping rates and equilibrium states. 

2PI-resummed effective actions and evolution equations are renormalisable \cite{Ren1, Ren2,Ren3,Ren4,Ren5,Ren6,Ren7}, but  at  a  substantial  numerical  cost. Hence  for  the  majority  of  simulations  in  the  literature, one introduces a counterterm only for the mass \cite{Salle:2000hd,Arrizabalaga:2004iw,App4} to eliminate the leading dependence on UV cutoff variation, leaving only some log-divergent contributions. We will do likewise, and to obtain a renormalized mass $am$ in our simulations, we need to introduce the bare mass parameter $am_0$ by
\begin{align}
    (am_0)^2  = (am)^2 - \lambda C \frac{1}{N_x^3} \sum_{k_{\rm lat}} \frac{1}{2 \sqrt{(ak_{\rm lat})^2 + (am)^2}}
\end{align}
The $N$-dependent prefactor $C$ depends on the expansion and the order, since the mass contribution are different at LO. We take $C=\frac{N+2}{6N}$ (loop expansion at LO and both at NLO), and $C=\frac{1}{6}$ ($1/N$-expansion at LO).

\section{Comparison of loop and 1/N expansion}
\label{sec:CompLaN}

\begin{figure}[ht]
  \includegraphics[width=0.5\textwidth]{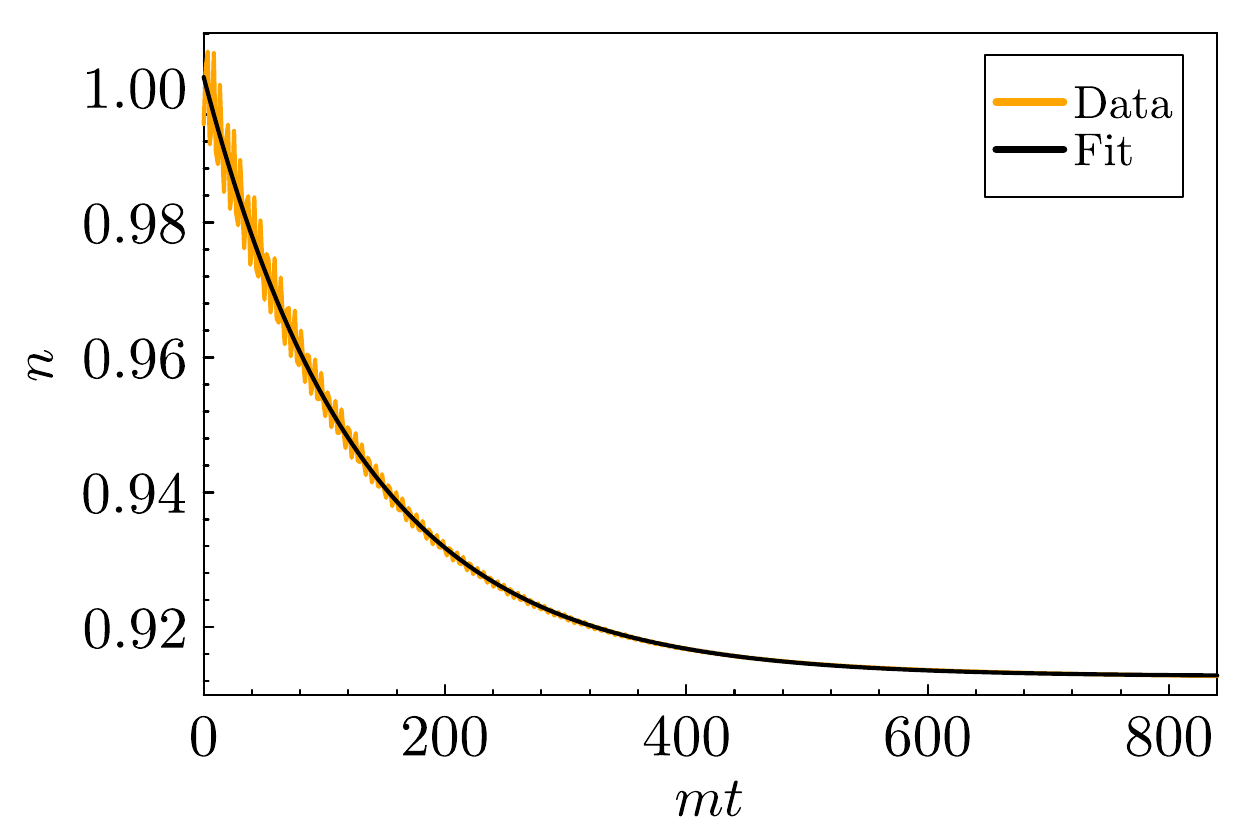}
  \includegraphics[width=0.5\textwidth]{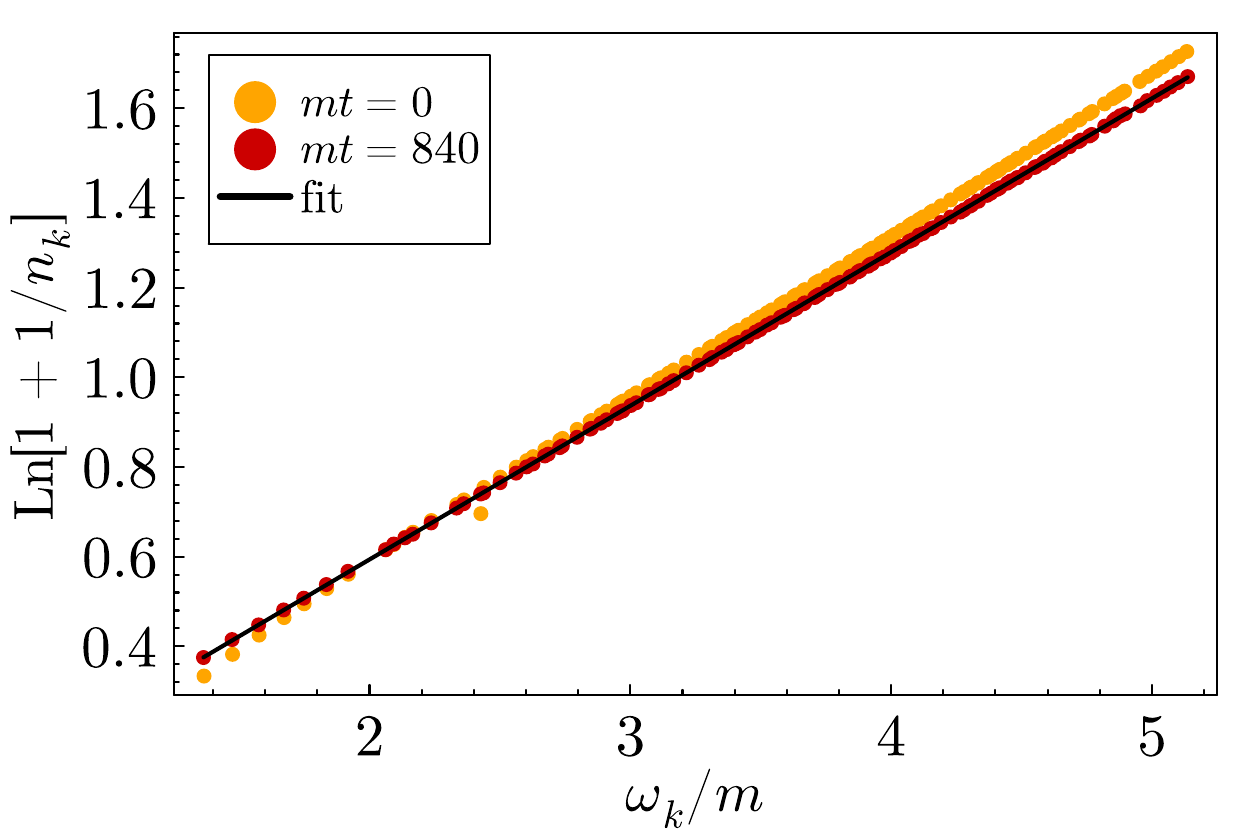}
  \caption{Left: Evolution of the initially plucked mode and the corresponding fit. The mode is initially displaced by $10\%$. Right: Initial and final state of a simulation. The initially free thermal state is brought to the thermal state of the interacting theory. For both, the initial temperature is $T/m=3$.}
  \label{fig:ExpExampleSimulation}
\end{figure}

The two diagram expansions are expected to approximate the exact result in each their limit of validity. The loop expansion naively when $\lambda$ is small, although it resums an infinite subset of diagrams at all orders in $\lambda$, while neglecting others. The $1/N$ expansion reorganises and further resums diagrams at the same order in $1/N$, but can only be expected to be a good approximation when $N$ is large. Large in this context has been found to be $N=10$ or even $20$ \cite{Aarts:2008wz,Aarts:2006cv}, while simulations are often done at $N=4$, inspired by the 4 real fields of the Standard Model Higgs field. For many applications in cosmology, $N=1,2$. One may also ask whether when $N$ is large enough that NLO approximates the exact result, it is also so large that it is effectively LO.

In the following we will compare the $\lambda$ and $1/N$ expansions at NLO, through computing the near-equilibrium damping rate of a single excited momentum mode. The initial propagators are initialized as a free thermal quantum system with temperature $T$, and the self-consistent thermal mass. On top of this near-equilibrium state, the mode $k_L/m = 2$ is "plucked" (like a string) by adding $10 \%$ to the occupation number. 
The initial thermal mass is obtained by solving   
\begin{equation}
    M_{\rm gap}^2 = m^2 + \lambda\frac{N+2}{6N}\int \frac{d^3k}{(2\pi)^3}\frac{n_k(T) + 1/2}{\sqrt{k_L^2 + M_{\rm gap}^2}}.
\label{eq:gapmass}
\end{equation}

Figure \ref{fig:ExpExampleSimulation} (left) shows such a  plucked mode approaching its equilibrium value, as well as the best fit of the form
\begin{equation}
    n(t) = n_{\rm eq}+(n_{\rm init}-n_{\rm eq})e^{-\gamma t}.
    \label{eq:nplunkfit}
\end{equation}
While the excited mode relaxes, the time evolution also adjusts the system from its initial free field state to the thermal state of the interacting theory. Since we initialize close to equilibrium it does not require a lot of energy redistribution between modes. In Figure \ref{fig:ExpExampleSimulation} (right) we show the initial and the final spectrum. The slope corresponds to the inverse temperature and the offset at $\omega_k=0$ to a chemical potential. The figure also shows a fit to the relation  
\begin{equation}
    \ln \big(1 + 1/n_k \big) = (\omega_k-\mu_{ch})/T,
\label{eq:BEdistlinear}
\end{equation}
where $\omega_k$ is the derived dispersion relation (\ref{eq:disprelation}). The true chemical potential must approach zero for long simulations since particle numbers are not protected by any symmetry. However, for larger values of $\lambda$, the deviation from a pure Bose-Einstein distribution appears like a non-zero late-time chemical potential.

\subsection{Large N results}
\label{sec:LargeNresults}

\begin{figure}[ht]
  \includegraphics[width=0.5\textwidth]{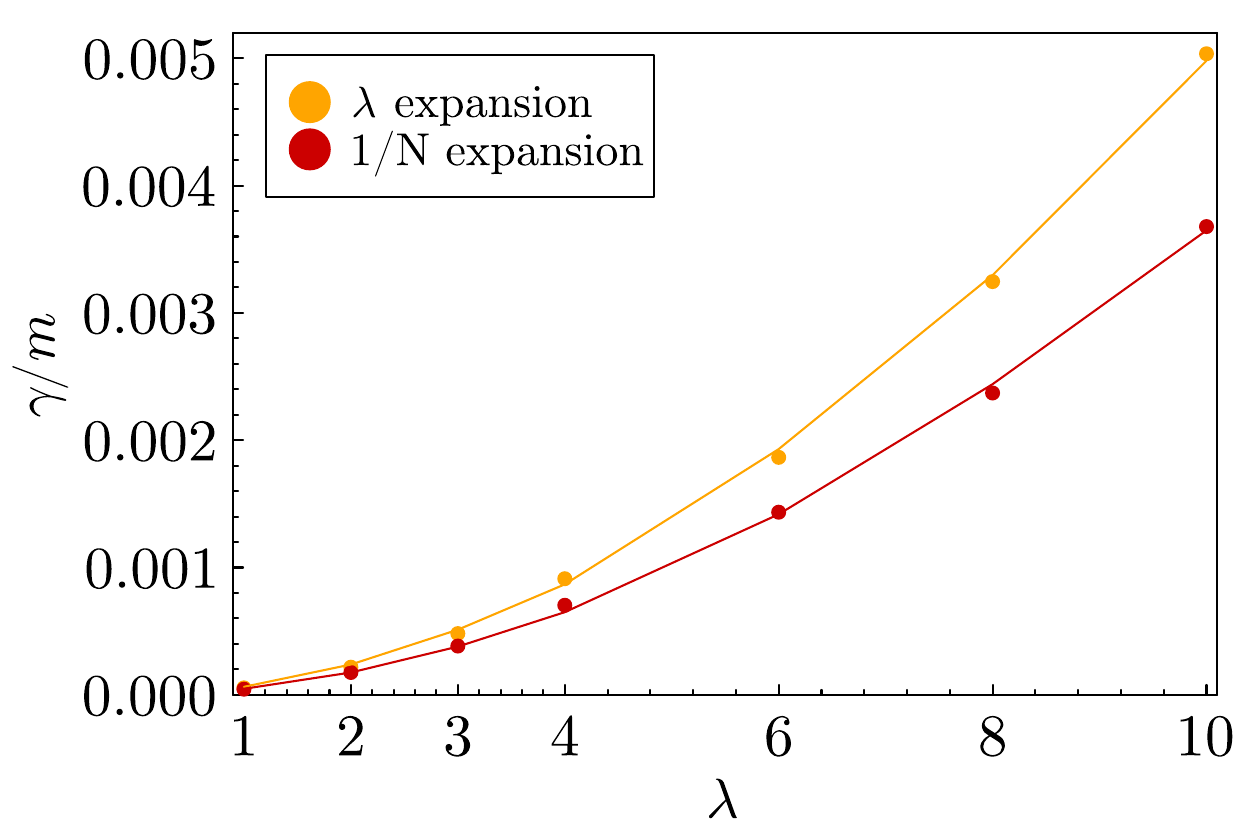}
  \includegraphics[width=0.5\textwidth]{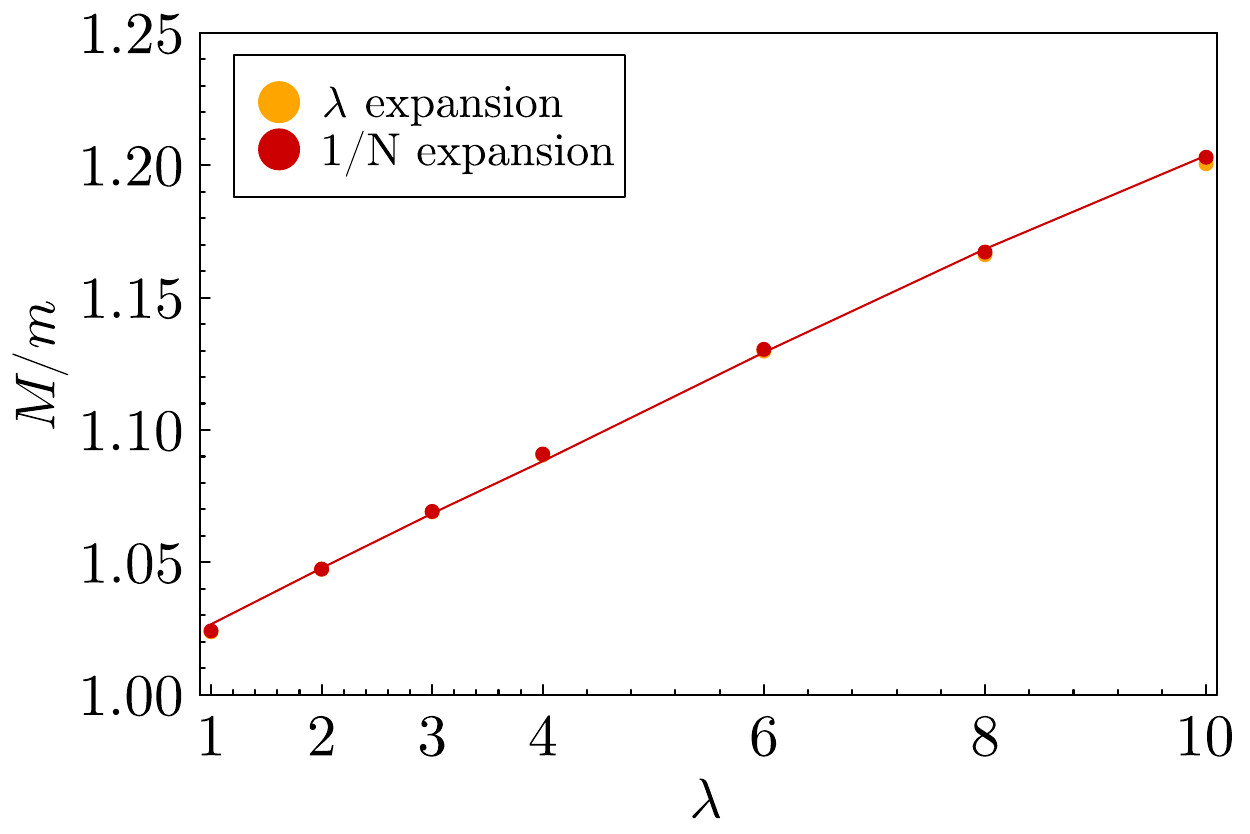}
  \caption{Left: The damping rate of the plucked mode for $N=10$ and various values of $\lambda$. The $1/N$ expansion is expected to be accurate for all values of $\lambda$. Right: The Hartree mass for both expansions agree very well. Both plots are obtained by initializing with $T/m=2$.}
  \label{fig:ExpT2k2varL_DampingrateMass}
\end{figure}

\begin{figure}[ht]
  \includegraphics[width=0.5\textwidth]{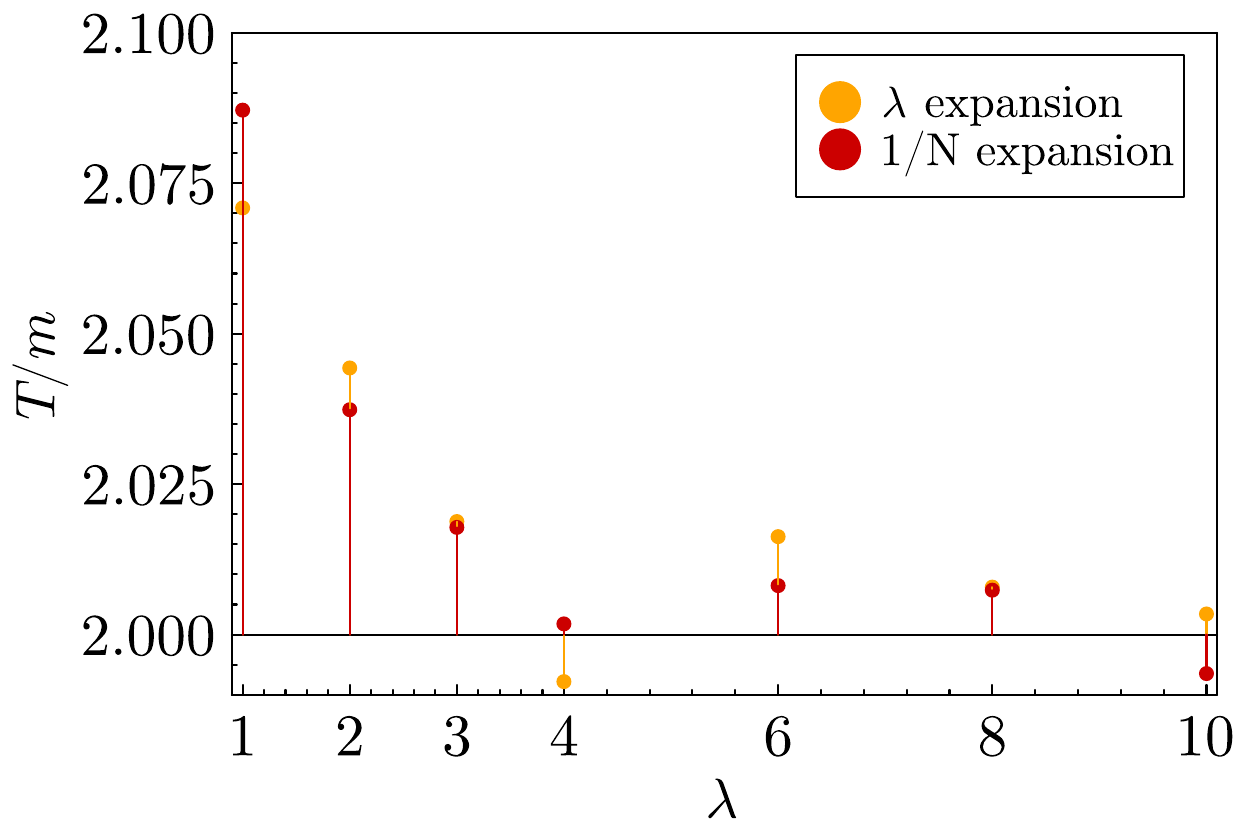}
  \includegraphics[width=0.5\textwidth]{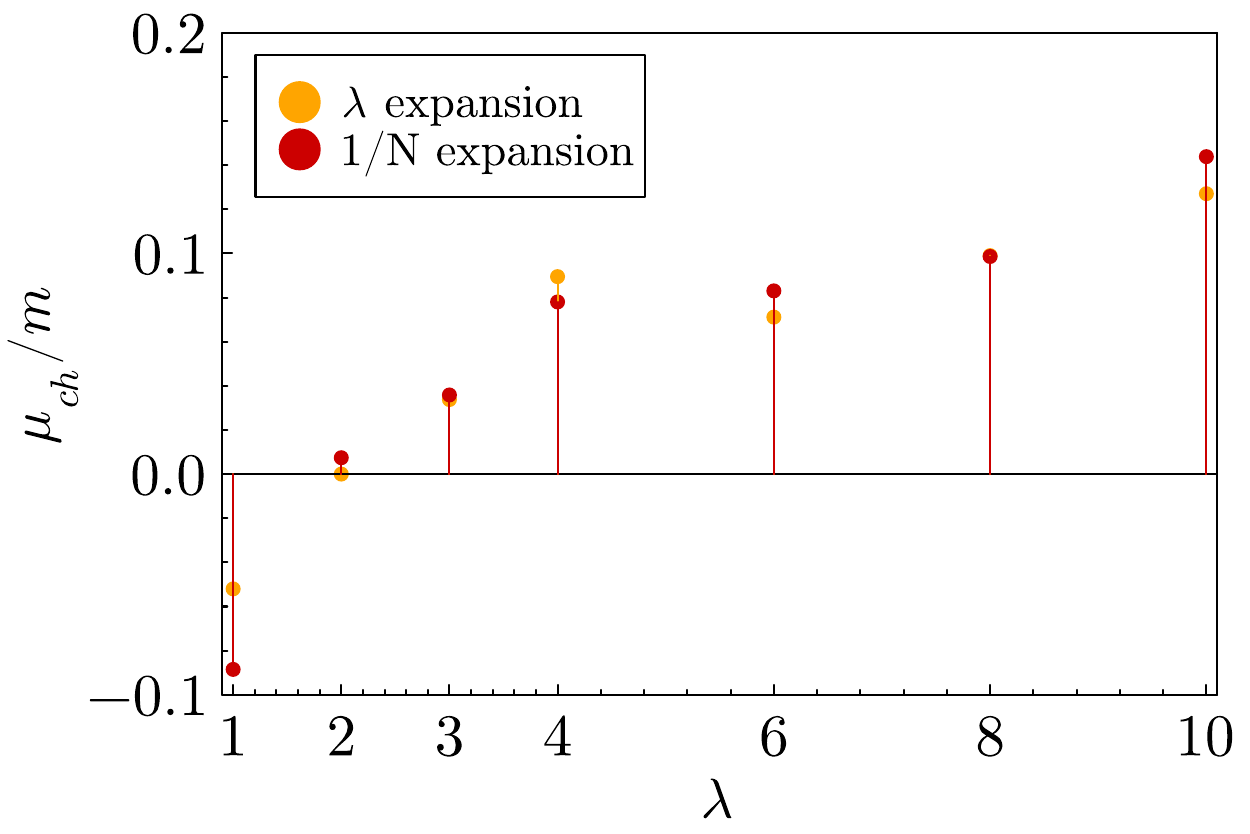}
  \caption{Temperatures (left) and effective chemical potentials (right) at the end of the simulations. Both plots are obtained by setting $N=10$ and $T/m=2$.}
  \label{fig:ExpT2k2varL_finalState}
\end{figure}
Figure \ref{fig:ExpT2k2varL_DampingrateMass} (left) shows the damping rate of the plucked mode at NLO for the $\lambda$ and $1/N$ expansions, respectively. We take a fairly large $N=10$, and we interpret the results as probing the applicability of the $\lambda$ expansion for various values of $\lambda$. We see that the coupling expansion overestimates the damping rate for $\lambda>2$ ($\lambda/N=0.2$), with a discrepancy of about $30\%$ at $\lambda=10$ ($\lambda/N=1$).

On general grounds (see also \cite{Wang:1995qf,Parwani:1991gq}), for weak coupling and high temperatures one would expect a parametric dependence of the damping rate of the form
\begin{equation}
    \gamma \propto \frac{ \lambda^2 T^2}{M},
    \label{eq:gammazeromodelambda}
\end{equation}
where $M$ is the effective mass of the system. 
Figure \ref{fig:ExpT2k2varL_DampingrateMass} (left) also includes fits of this form, which agree well with the simulations. We perform the fit inserting the measured temperatures and the Hartree effective mass values, extracted from the local self-energy diagrams and shown on the right of Figure \ref{fig:ExpT2k2varL_DampingrateMass}. Overlaid is a fit of the form
\begin{equation}
    M^2 - m^2 \propto \lambda T^2,
    \label{eq:Hartreefit}
\end{equation}
where T refers again to the measured temperature. Figure \ref{fig:ExpT2k2varL_finalState} shows this effective $T$ and $\mu_{ch}$ at the end of the simulation. For large coupling, the deviation from the initial free field thermal state increases, but the final temperature is still within $5 \%$ of the initial value. Similarly, non-zero coupling in effect produces a non-vanishing chemical potentials that is shown on the right of Figure \ref{fig:ExpT2k2varL_finalState}. The final state has little dependence on the diagram expansion. 

\subsection{Small $\lambda$ results}

\begin{figure}[ht]
  \includegraphics[width=0.5\textwidth]{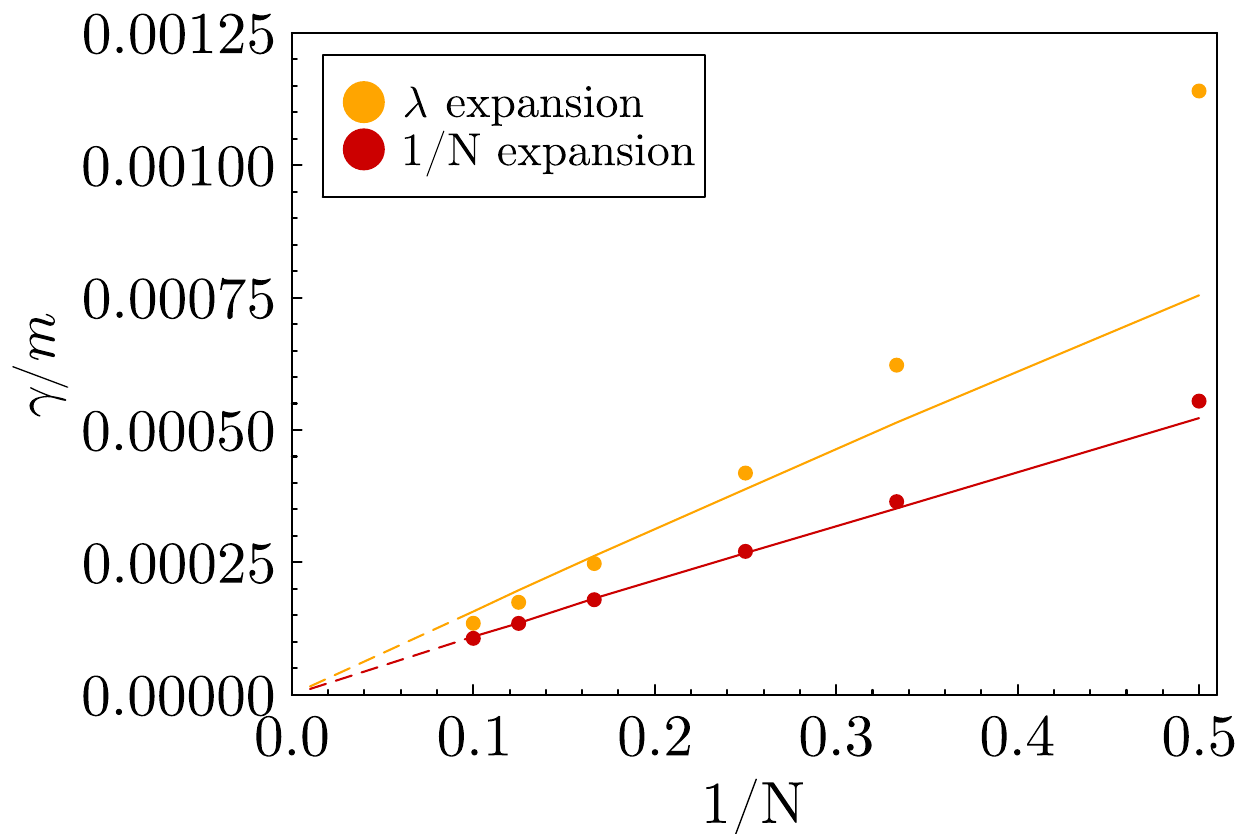}
  \includegraphics[width=0.5\textwidth]{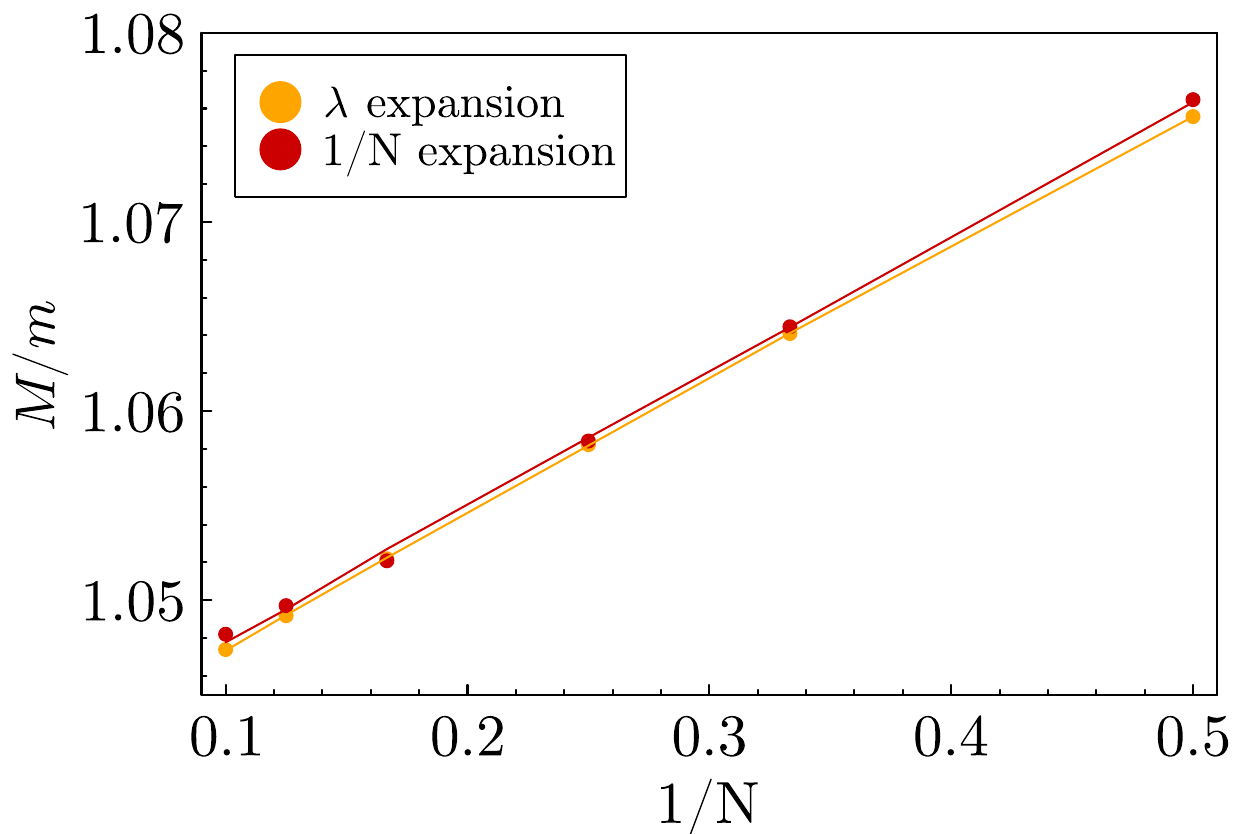}
  \caption{Left: The damping rate of the plucked mode for $\lambda=1$ and various values of $1/N$. Right: The effective mass for both expansions agree very well. Both plots are obtained by setting $T/m=3$.}
  \label{fig:ExpT2k2varN_DampingrateMass}
\end{figure}
We now set $\lambda$ to a small value of $1$, for which we expect the coupling expansion to be fairly reliable (although see below), and compute the damping rates for different values of $1/N$, as shown in Figure \ref{fig:ExpT2k2varN_DampingrateMass} (left). Naively, one may expect a dependence of the form
\begin{equation}
    \gamma \propto \frac{ T^2}{N M},
    \label{eq:gammazeromodeNinverse}
\end{equation}
and this linear dependence matches the $1/N$ expansion. The coupling expansion damping rate is however not linear in $1/N$, and is consistently larger, with a factor of two for $N=2$. 
In Figure \ref{fig:ExpT2k2varN_DampingrateMass} (left) we show the corresponding Hartree masses. We overlay the fit
\begin{equation}
    M^2 - m^2 \propto \frac{\lambda T^2}{N}.
    \label{eq:HartreefitNinverse}
\end{equation}

\subsection{$T$-dependence}
\label{sec:Tdep}

\begin{figure}[ht]
  \includegraphics[width=0.5\textwidth]{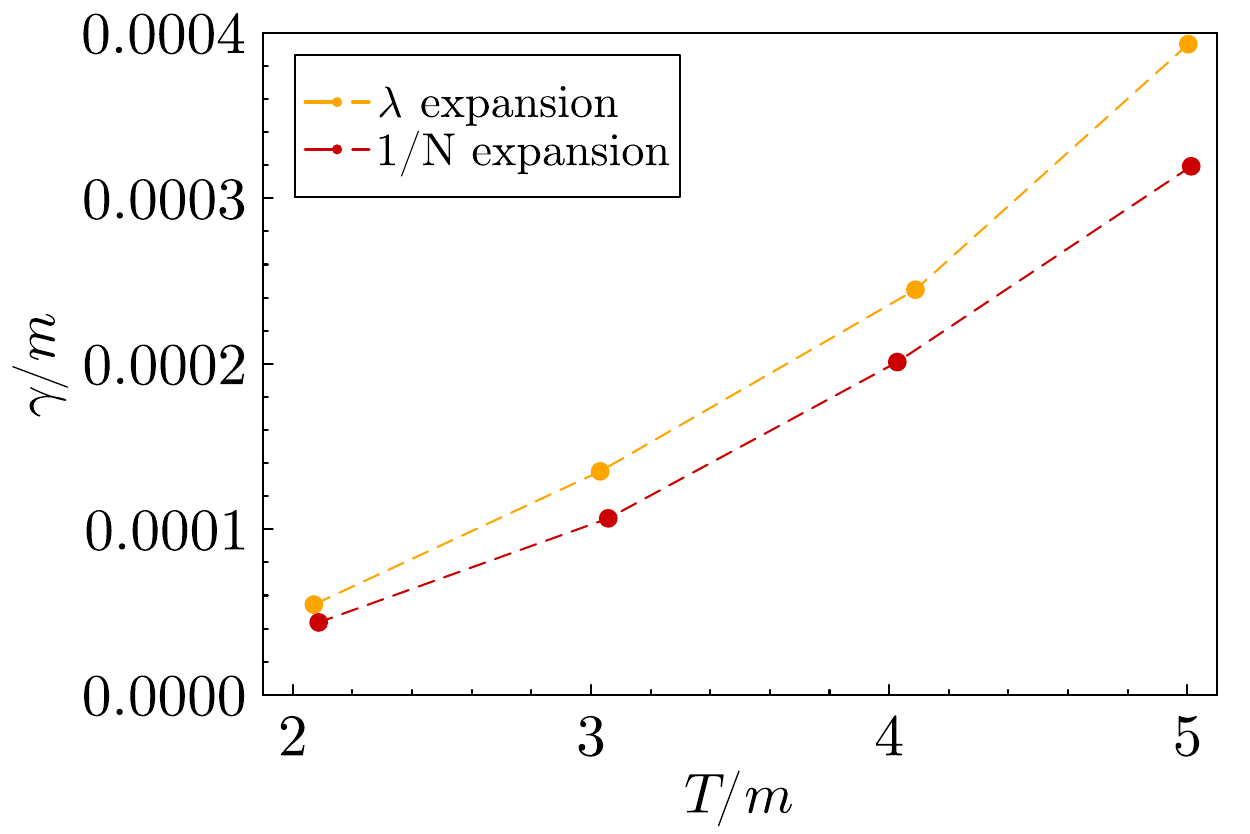}
  \includegraphics[width=0.5\textwidth]{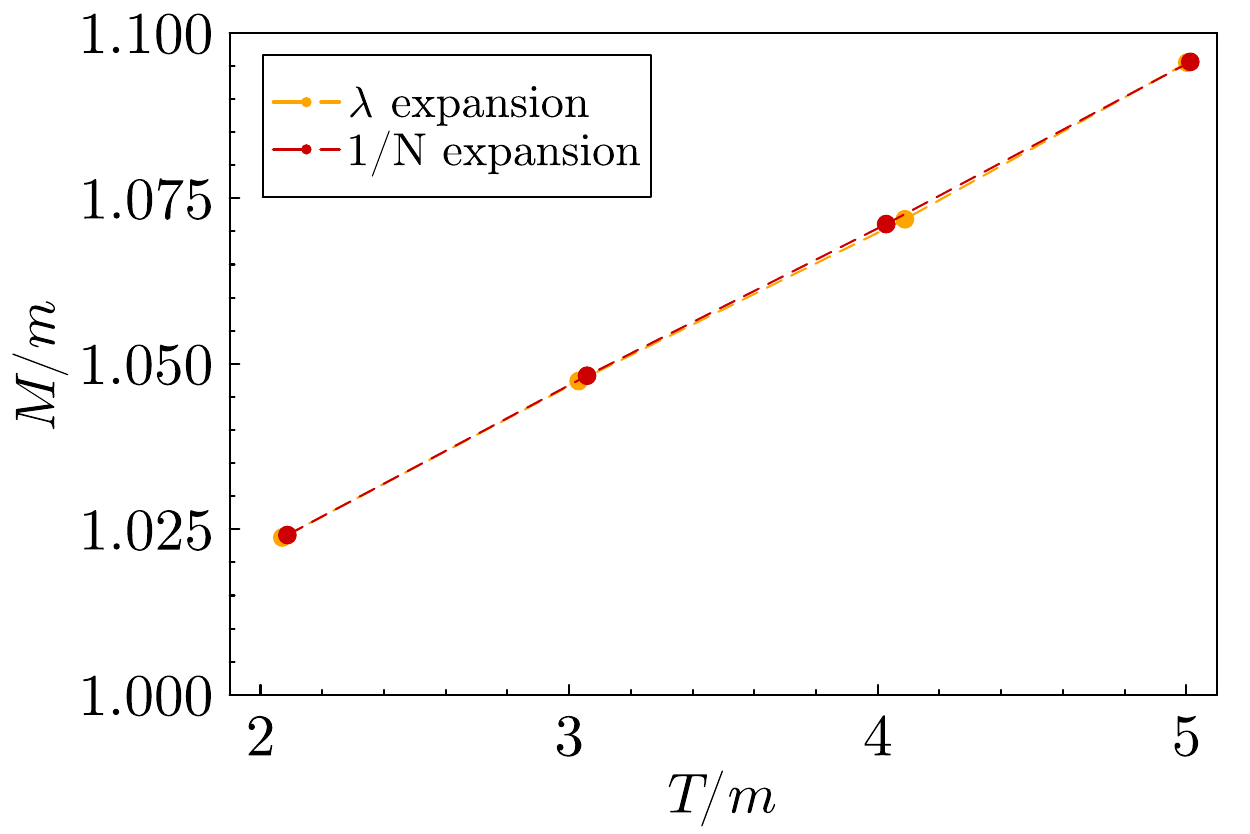}
  \caption{The damping rate (left) and the effective mass (right) for different values of $T/m$. Parameters are $\lambda=1$ and $N=10$.}
  \label{fig:Expk2varT}
\end{figure}

In order to complete the numerically quite challenging simulations, the damping rate should not be too small. This implies, that we cannot simultaneously have small coupling, large $N$ and small temperature $T$. As a last check in this chapter we analyse the temperature dependence of the obtained results. 

We first note that for large values of temperature the discretization effects become relevant. To see this we calculate the occupation number for the highest available mode on the lattice\footnote{The lattice momentum values are given by $(a k)_{L}^2 = \sum_{i=1}^3 \left(2 - 2 \cos  k_i \right)$, $ k_i = \frac{2 \pi}{N_x} n_i$, $n_i \in \Big(-\frac{N_x}{2}+1, \frac{N_x}{2}\Big)$.} which is given by
\begin{align}
   (k_{L}/m)_{max}^2 = 12 / (am)^2.
\end{align}
The particle number for this mode is given by
\begin{equation}
    n_{k_L} = \frac{1}{e^{\omega_k/T} -1 } = \frac{1}{e^{ \sqrt{1 + (k_{L}/m)_{max}^2}/(T/m)} - 1 }.
    \label{eq:HartreefitNinverse}
\end{equation}
 With a lattice spacing of $am=0.7$ we obtain $n_{k_L} = 0.09$ for $T/m=2$, $n_{k_L} = 0.23$ for $T/m=3$. Clearly, for large temperatures the spectrum is cut off in the UV by the discretization rather than the physical exponential suppression.

We show in Figure \ref{fig:Expk2varT} the damping rate with $\lambda=1$ and $N=10$, where both expansions should give similarly valid approximations. We see that the agreement improves for smaller values of $T/m$, where more of the spectrum is resolved by the lattice. 
    
\section{Classical limit of 2PI evolution and the classical-statistical approximation}
\label{sec:ClassComp}

\begin{figure}[ht]
  \includegraphics[width=0.5\textwidth]{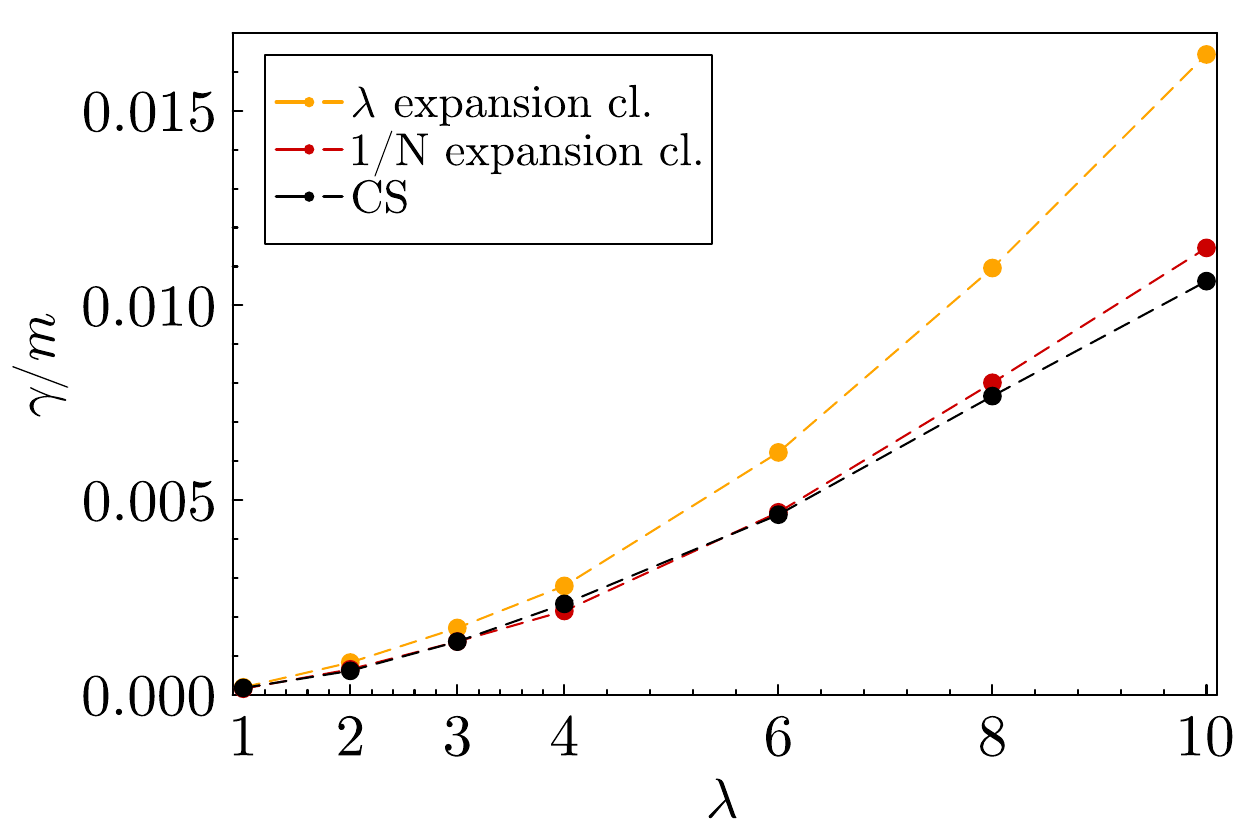}
  \includegraphics[width=0.5\textwidth]{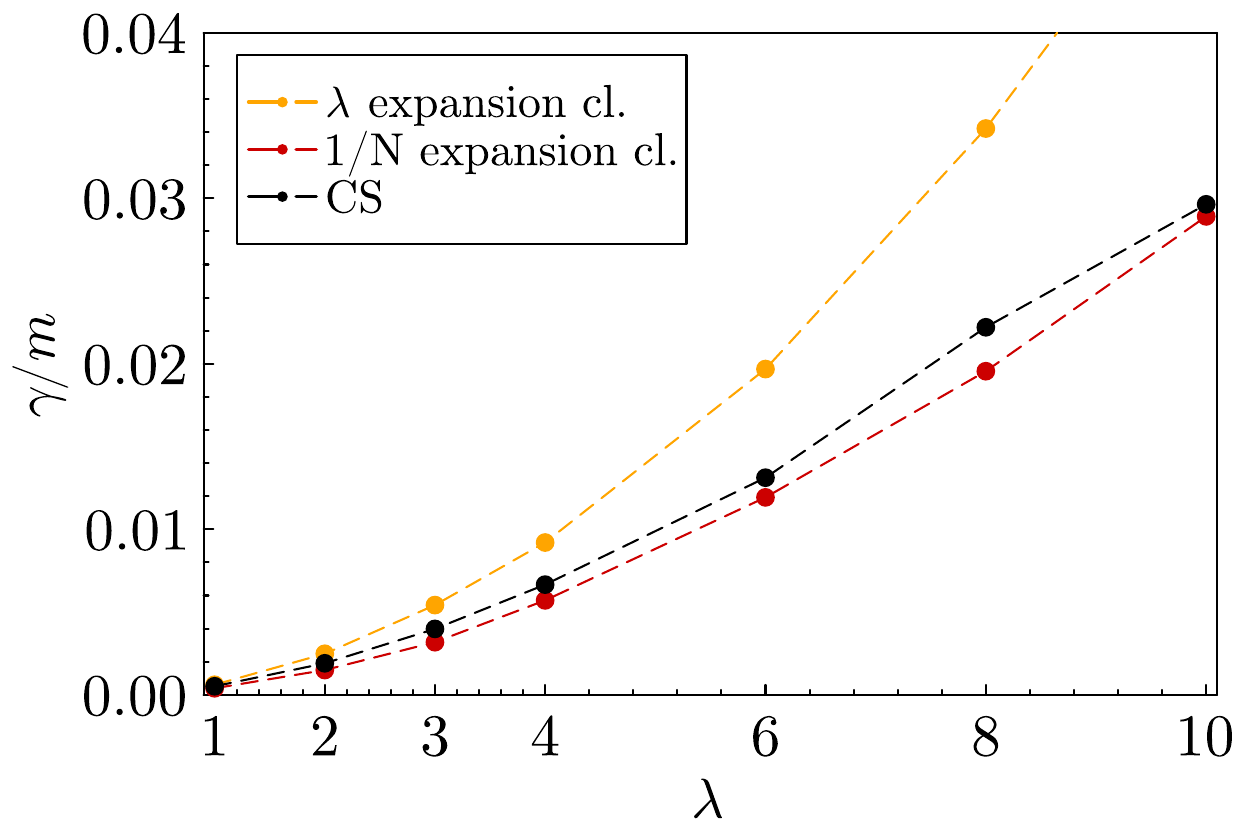}
  \caption{Left: The damping rate in the classical limit of the 2PI evolution and the exact result obtained from classical-statistical simulations for $N=10$. Right: Same as the left-hand figure, but for $N=4$. Both plots are obtained by setting $T/m=3$.}
  \label{fig:Clak2varL}
\end{figure}

\begin{figure}[ht]
  \centering
  \includegraphics[width=0.5\textwidth]{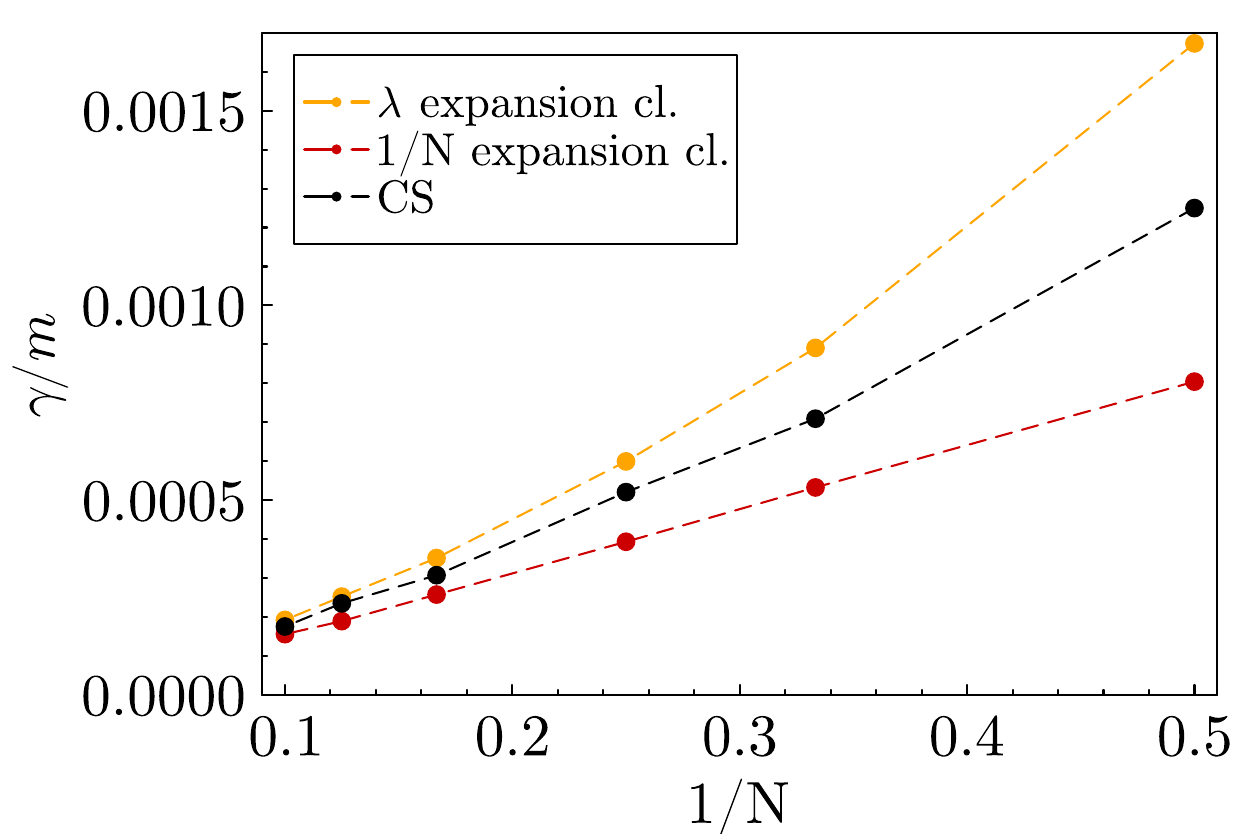}
  \caption{Damping rate in the classical limit of the 2PI evolution and the exact result obtained from classical-statistical simulations for $\lambda=1$ and $T/m=3$.}
  \label{fig:Clak2varN}
\end{figure}

As discussed in the previous chapter we have good reason to believe in the applicability of the $1/N$ expansion for $N=10$ for all reasonable values of $\lambda$. For the case of small $\lambda$, but different $N$ the case is less clear. Ideally, one would proceed to NNLO or to the full theory to compare, but unfortunately this is extremely challenging for the full quantum theory (although see \cite{Aarts:2006cv}). What we can do is to take the classical limit of the 2PI evolution equations and compare to a classical-statistical computation  \cite{Aarts:2001yn}, which for a classical theory is exact.\footnote{One could also compare the classical-statistical simulation as an approximation to the quantum result, but this has the caveat that the background state is then in the process of thermalising to the (cut-off dependent) classical equilibrium state.} 

The classical-statistical approach is to generate an ensemble of initial conditions representing a given initial state, evolve them using the classical equations of motion, and compute observables as simple averages over the realisations in the ensemble. This is a numerically straightforward procedure, applicable to explicitly classical systems, or as an approximation to quantum systems, when the occupation numbers are large $n_k\gg 1$ (see for instance \cite{Smit:2002yg,Garcia-Bellido:2002fsq}). In our simulation here, we use the exact same lattice discretisation as for the 2PI evolution, and average over an ensemble of 5000 realisations. This is sufficient for the statistical errors to be under control, so that the averaged time evolution of the plunked mode is smooth and well described by the exponential fit in eq. (\ref{eq:nplunkfit}).    

The 2PI evolution equations also have a classical limit, which amounts to discarding instances of $\rho^2$ (or $\rho I_\rho$) when appearing in the combination $F^2-\rho^2$ (or $FI_F-\rho I_\rho$), in eqs. (\ref{eq:NLOlambda}, \ref{eq:NLON}, \ref{eq:NLONI}). This can be made more rigorous in terms of discarding a vertex and a set of diagrams in the Keldysh field basis \cite{Aarts:1997kp,Millington:2020vkg}, and is consistent with the requirement $n_k\gg 1$, since $F\simeq n_k$ and $\rho\simeq 1$. 

We note that the classical evolution does not distinguish between zero-point fluctuations and other excitiations. In this and the following chapter we keep the definition of $n_k$
in eq. \ref{eq:particlenrdefinition}, but keep in mind that the classical occupation number $\tilde{n}_k = n_k +1/2$ additionally includes a half in each mode.

Figure \ref{fig:Clak2varL} shows the damping rate for $N=10$ (left) and $N=4$ (right), varying $\lambda$. The first is the classical limit of figure \ref{fig:ExpT2k2varL_DampingrateMass} (left). 
We see that for $N=10$ the $1/N$ truncation at NLO indeed is a very good approximation to the exact result. But we also see that even at $N=4$ the agreement is fairly good. For both values of $N$ the accuracy is within $10\%$ of the exact result. The accuracy of the loop expansion systematically worsens for larger coupling, and exceeds $50\%$ at $\lambda=10$.
Figure \ref{fig:Clak2varN} shows the same for various values of $1/N$, the classical analogue of figure \ref{fig:ExpT2k2varL_DampingrateMass} (left). We see that the situation is indeed less clear. It seems that the exact result of the classical statistical simulation sits half-way between the $\lambda$ and $1/N$ expansion results. Apparently, $\lambda=1$ is not that small after all, and even at $N=2$, the $1/N$-NLO at $N=2$ is not worse than $\lambda$-NLO at this coupling (see also \cite{Aarts:2001yn}).

\section{Classical thermalisation}
\label{sec:ClasTherm}

\begin{figure}[ht]
  \includegraphics[width=0.5\textwidth]{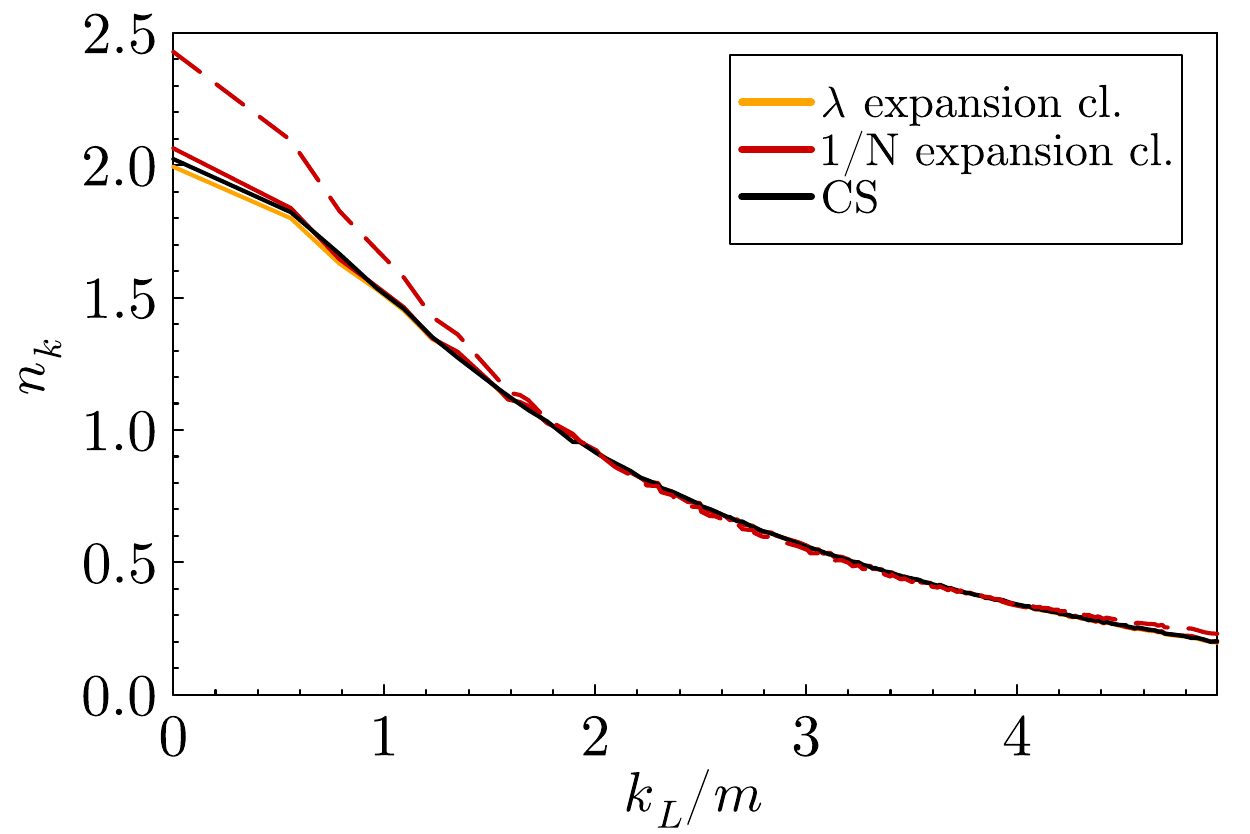}
  \includegraphics[width=0.5\textwidth]{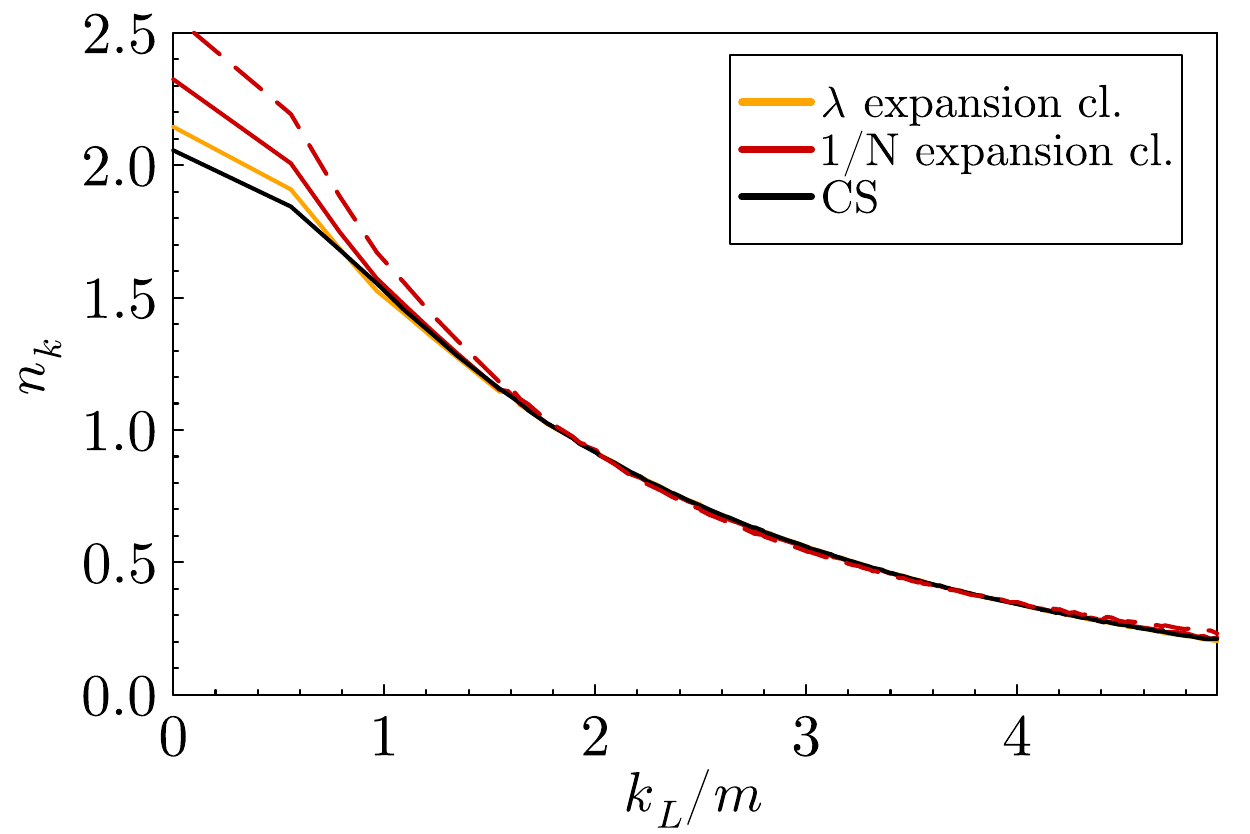}
  \caption{Left: Comparison of the final near-equilibrium state in the classical limit of 2PI truncations and the classical-statistical approximation. The plots show also the quantum final state in the $1/N$ expansion. $N=2$ (left) and $N=10$ (right). Both plots are obtained by setting $\lambda=1$, $T/m=3$.}
  \label{fig:Clak2varx_finalstate}
\end{figure}

\begin{figure}[ht]
  \centering
  \includegraphics[width=0.5\textwidth]{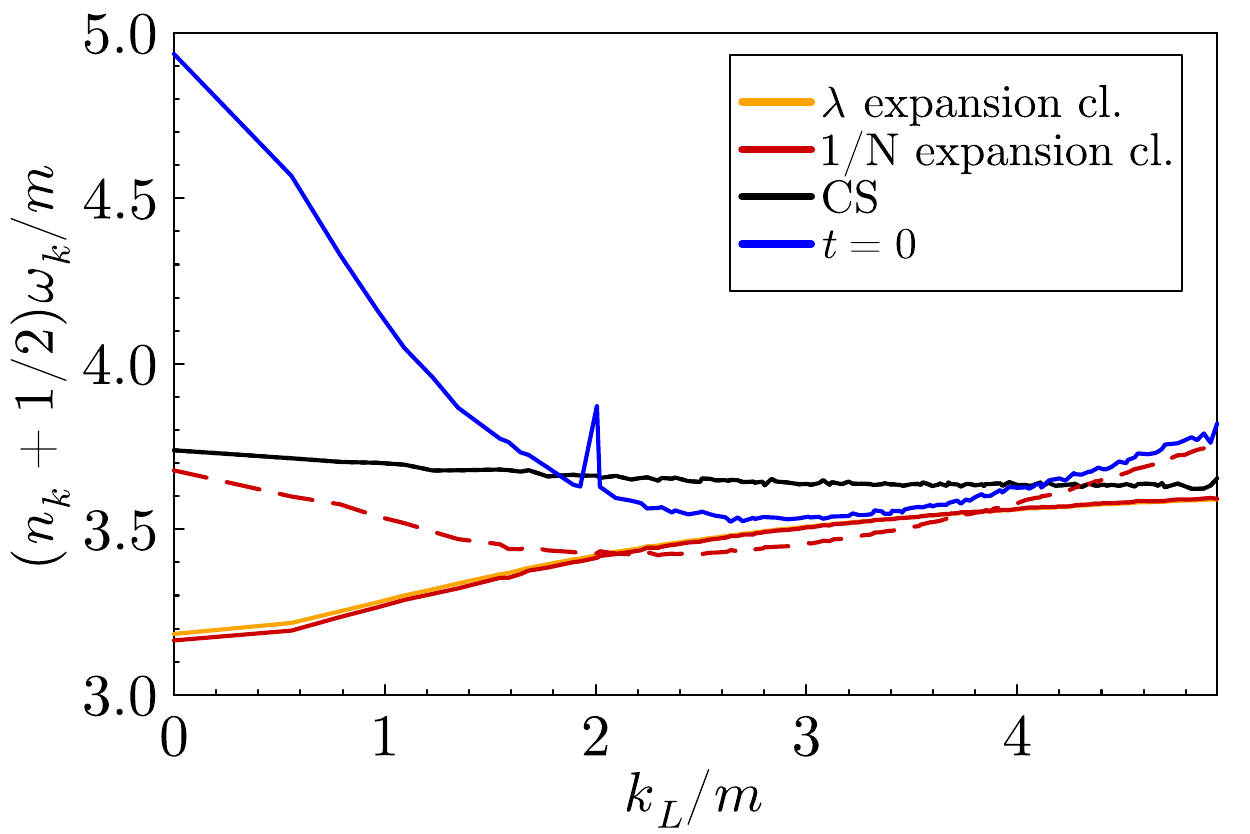}
  \caption{Comparison of $(n_k+1/2) \omega_k$ near at the near-equilibrium state in the classical limit of 2PI truncations and the classical-statistical approximation. The plots show also the initial state and the final state of the quantum evolution in the $1/N$ expansion. Plot is obtained by setting $N=10$, $\lambda=10$, $T/m=3$.}
  \label{fig:Clak2varx_finalpipistate}
\end{figure}

The equilibrium state of classical simulations is not a Bose-Einstein distribution, it is cut-off dependent and therefore no sensible physical temperature can in principle be extracted \cite{Epelbaum:2014yja}. On a finite lattice, the system is however expected to approach classical equipartition, where by $\omega_k n_k=T$, independent of $k$.

Figure \ref{fig:Clak2varx_finalstate} shows the late-time spectrum in the classical limit, for $N=2$ (left) $N=10$ (right) and $\lambda=1$ and $T/m=3$. The spectra are very similar in the UV, but differ in the far IR for $N=10$. We believe that this could be a finite volume effect. We also show the final state in the $1/N$ expansion in the quantum NLO truncation (dashed line), which differs for a broad range of momenta in the IR and in the far UV for both large and small $N$. While the quantum evolution ensures that zero-point fluctuations (the "1/2" in the $n_k+1/2$) remain in all modes, the classical evolution equation do not make such a distinction, and will redistribute all fluctuations into a classical equilibrium state. For our present purposes of comparing like for like, we have initialised also the classical simulations including zero-point fluctuations, but this is in general problematic, since it introduces a divergent (cut-off dependent) energy density into the system (see also \cite{Tranberg:2022noe}).

Figure \ref{fig:Clak2varx_finalpipistate} shows $(n_k + 1/2) \omega_k$ in the classical limit at late times for $N=10$ and $\lambda=10$. It is of interest since in the classical limit it is constant as the classical particle number is $(n_k + 1/2) \approx T/\omega_k$. The classical evolution indeed shows this constant value. Both 2PI truncations in the classical limit show a surprisingly similar behaviour, whose discrepancy to the true solution increases in the IR. The plot also shows $(n_k + 1/2) \omega_k$ of the initial distribution (blue) and at late-stages of the quantum evolution in the $1/N$ truncation. 

\section{Discretisation schemes}
\label{sec:Discret}

\begin{figure}[ht]
  \includegraphics[width=0.5\textwidth]{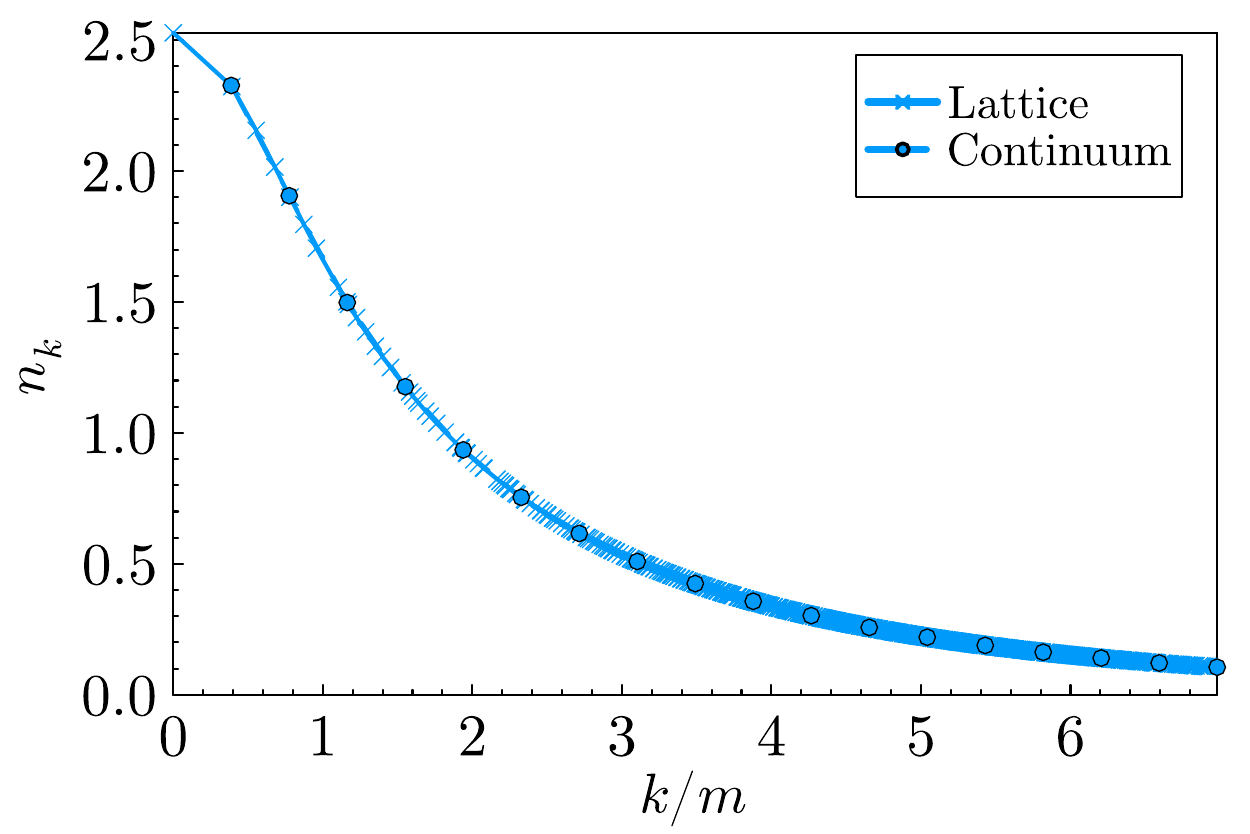}
  \includegraphics[width=0.5\textwidth]{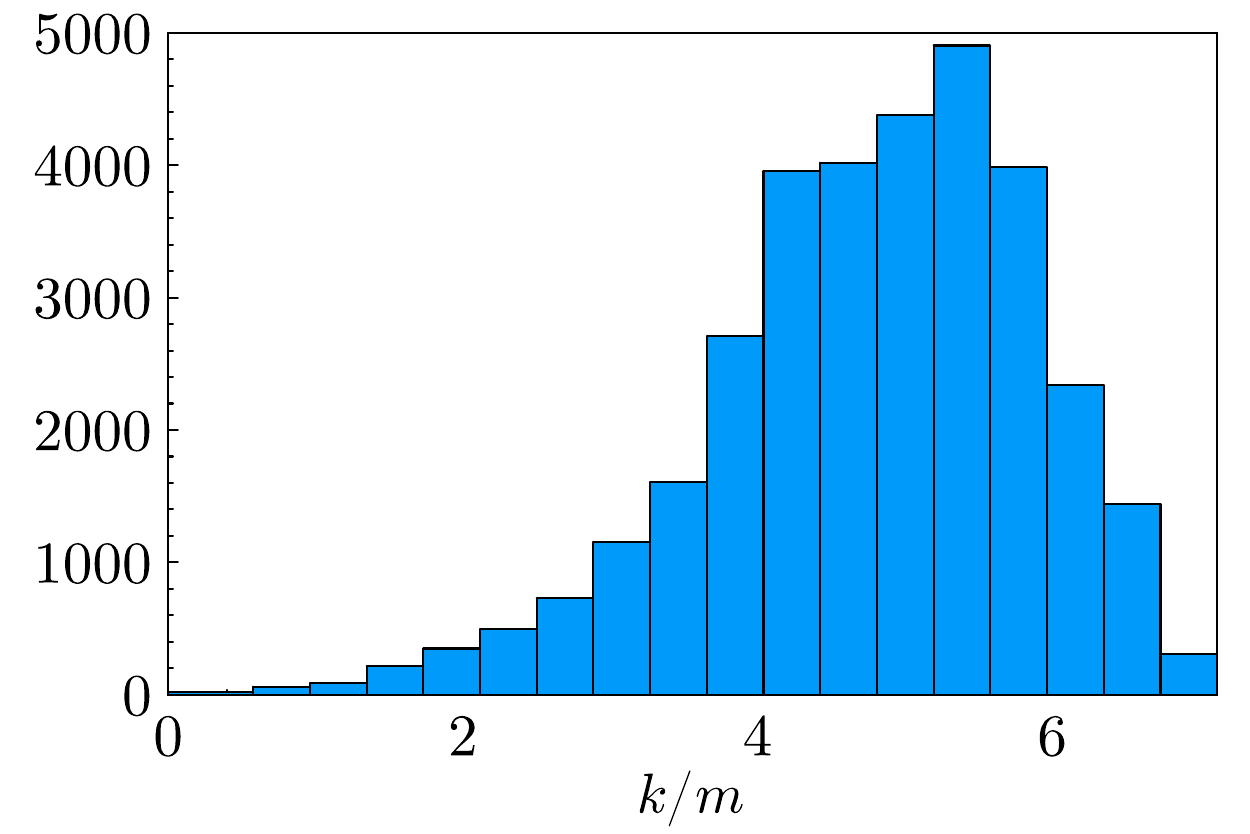}
  \caption{Left: Bose-Einstein particle distribution with $T/m=3$ in the two discretisation schemes. Right: Lattice momenta binned according to the continuum momentum intervals. Plots are obtained by using the same IR and UV cut off values in both schemes according to eq. \ref{eq:sameDisc}.}
  \label{fig:CompDisc}
\end{figure}
Up to this point, we have discretised the system on a space-time lattice at the level of the action, as described in section \ref{sec:IniConds}, using $N_x^3 \times N_t$ lattice points. 
The available momenta are discrete and related to the lattice spacing through $(ak)^2_{lat}= \sum_{i=1}^3 (2 - 2 \cos k_i)$, where $k_i = \frac{2 \pi}{N_x} n_i$ and $n_i \in (-\frac{N_x}{2} + 1, \frac{N_x}{2})$. This standard discretization inherits all the (dis)advantages of lattice field theory (symmetries, continuum limit, renormalisability, gauge symmetry, fermion doublers). For the present context, however, the majority of the computational cost is spent in the UV since the lattice modes bunch up there. Secondly, since the Fourier components of a spatial homogeneous system $F(x,y) = F(t,t^\prime,x-y)$ are functions of momenta amplitudes $|k|$ only, $F_k(t, t^{\prime})$, many of the reciprocal lattice sites $(k_x,k_y,k_z)$ contain duplicate information. 

Alternatively, the spatially homogeneous system may be studied using another discretisation scheme (see \cite{App17} and references therein), where the continuum equations of motions are solved through discretising an interval in momentum space into a finite number of equidistant values of $|k|$. The momenta are given by $(a k) = \frac{\pi}{D} (i+1)$, where $i \in (0,D-1)$. One notes the absence of the zero momentum node in this description, on the other hand there is no bunching up of modes in the UV, and all the modes have distinct information. From the point of view of renormalisation and Lorentz and gauge symmetry, this is a brute force momentum cut-off regularization.  

In both discretisation schemes, the UV cutoff $\Lambda_{UV}$ is given by the largest available momentum. In the lattice and continuum schemes it is given by 
\begin{align}
    (a \Lambda_{UV} )_{\rm lat} = \sqrt{12} \approx 3.4, \quad (a \Lambda_{UV} )_{\rm con} = \pi,
\end{align}
respectively.
In order to obtain the same UV cutoff in mass units the scales $am$ need to be related via 
\begin{align}
    \frac{\Lambda_{UV}}{m} = \frac{\sqrt{12}}{m_{\rm lat}} = \frac{\pi}{m_{\rm con}}.
\end{align}
Similarly, the IR cutoff is given by the smallest (non-zero) available momentum mode which additionally depends on the number of lattice sites:
\begin{align}
    \frac{\Lambda_{IR}}{m} = \frac{ 2 \pi}{N_x m_{\rm lat}} = \frac{\pi}{D m_{\rm con}}.
\end{align}
In order to compare the two schemes we analyse systems with the same IR and UV cutoff. Two equivalent discretisations are given by 
\begin{align}
   \text{Lattice: } & N_x = 32, \quad m_{\rm lat}  =0.5 \\ 
   \text{Continuum: } & D = 18, \quad m_{\rm cont} =0.45.
   \label{eq:sameDisc}
\end{align}
Figure \ref{fig:CompDisc} (left) shows the available momentum values in the two descriptions. While the continuum scheme features only $18$ momentum values, the lattice has $833$ distinct momenta where most are in the UV. On the right we additionally show the number of available lattice modes binned according to the $18$ momenta values in the continuum description. The total histogram count is $32^3$ \footnote{We note that all $32^3$ nodes are only used in the FFT algorithm to calculate the self-energy. The equation of motion is solved for $833$ distinct momenta values.}. 

\begin{figure}[ht]
  \includegraphics[width=0.5\textwidth]{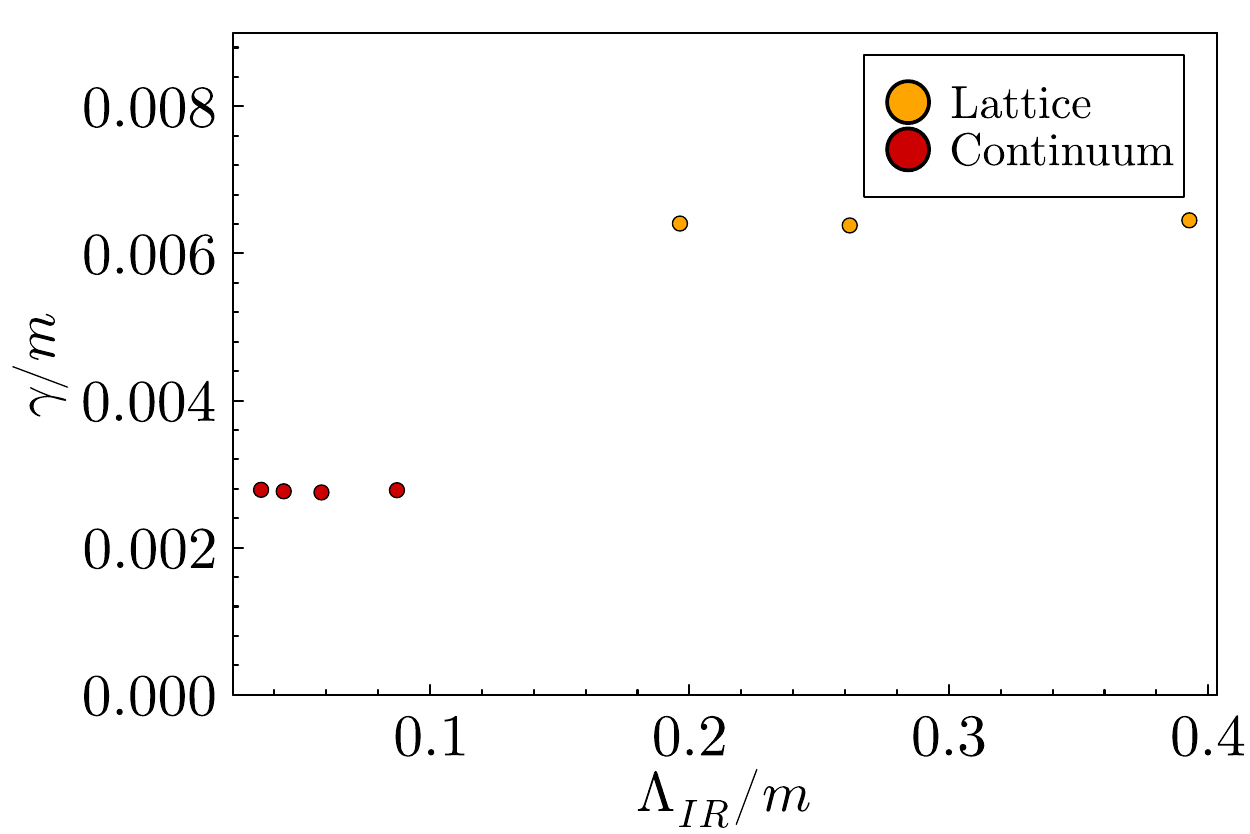}
  \includegraphics[width=0.5\textwidth]{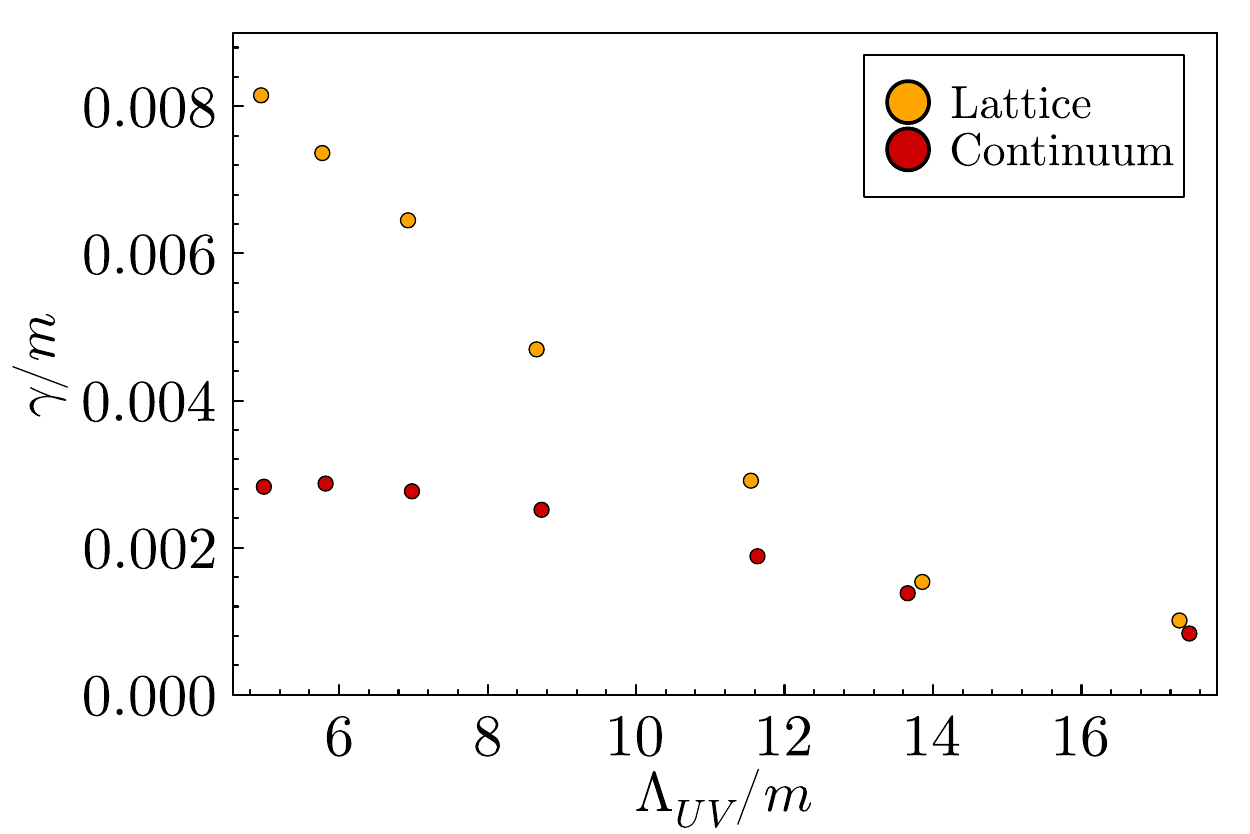}
  \caption{Left: Damping rates for various infrared cutoffs corresponding to $N_x=32,48,64$ and $D=80,120,160,180$. The ultraviolet cutoff is set at $\frac{\Lambda_{UV}}{m} = 6.9$. Right: Damping rates for various UV cut offs. We see that both discretisations agree well from $\frac{\Lambda_{UV}}{m} = 14$ onwards corresponding to $(am)_{lat}=0.2$. Plots are obtained in the $1/N$ expansion and by setting $\lambda=10$, $N=10$, $T/m=3$.}
  \label{fig:DiscDamping}
\end{figure}

In the following we increase the number of continuum modes to $160$, since $D=18$ modes is a very crude approximation. To reach the lowest-$k$ mode on the lattice, we would need $N_x\gg 32$ which is beyond our reach. Conveniently, as shown in Figure \ref{fig:DiscDamping} (left), already by $N_x=32$, the volume dependence is very small. 

In Figure \ref{fig:DiscDamping} (right), we show the familiar damping rate for various values for the UV cutoff. We see that the results for the damping rate differ by a factor of $2-3$ for low cutoffs. As the cut-off increases (lattice spacing decreases) the agreement improves and at around $\frac{\Lambda_{UV}}{m} = 14$ ($am=0.2$) agrees very well. 
In the present case the continuum discretisation saves a factor of about two in simulation time, which scales linearly with the number of continuum modes, $D$. 

\section{Conclusion}
\label{sec:Conclusion}

The 2PI effective action provides a compelling method for studying out-of-equilibrium processes, in particular in situations where the classical-statistical approximation is insufficient. The cost of including quantum effects is that it requires a diagrammatic expansion which must be truncated. The resulting equations are costly to integrate numerically, but still allow for the investigation of processes that occur on large time scales. In this work we focused on various aspects of the loop and $1/N$ expansion truncated at NLO. Through computing the relaxation of a displaced momentum mode from equilibrium, we were able to assess convergence properties of different approximations. 
We conclude that:
\begin{itemize}
\item In the classical limit, the $1/N$ expansion is the better approximation to the exact result, also at large coupling. The agreement is very good at $N=10$, but reasonable also in the popular, and physically more relevant, case of $N=4$. It should not be employed for $N=1,2$. The coupling expansion performs well for $\lambda < 2$ (for $N=10,4$). 
\item Also away from the classical limit, for the quantum evolution, the two expansions diverge for $\lambda >3$. To the extent that calibration in the classical limit translates away from this limit, this implies that one should be wary of the coupling expansion for larger couplings. 
\item In the classical limit, the 2PI-evolution indeed thermalises to the classical equilibrium, which is different from the quantum equilibrium in the IR (it is a classical spectrum, not a Bose-Einstein distribution) and in the UV (the classical spectrum is cut off by the momentum cut-off and does not distinguish zero point fluctuations from actual particle excitations). 
\item The lattice and continuum discretisation prescriptions, necessary for numerical implementation, give similar results for cut-offs corresponding to a lattice spacing of $am\lesssim 0.2$ (at least for the damping rate in a scalar field theory). For homogeneous systems, finite volume effects are already quite small for the lattice volumes considered here. 
\end{itemize}
 The numerical effort of 2PI simulations is nowadays manageable also in 3+1 dimensions and for late times. While it would certainly be useful to compare to other diagram expansions and/or the full NNLO, both coupling and $1/N$ truncations at NLO seem to be reliable for a reasonable range of parameters. 
 Next up is to resolve the issue of gauge fields, ideally in combination with scalar fields and/or fermions, and scaling simulations up to even larger physical volumes. Whether in time also inhomogeneous systems can be treated beyond LO remains to be seen. 
 
\subsubsection*{Acknowledgements}
We thank Alexander Rothkopf and Daniel Alvestad for pleasant discussions and comments. The numerical work was performed on resources provided by Sigma2 - the National Infrastructure for High-Performance Computing and Data Storage in Norway.

\bibliography{biblio.bib}
\bibliographystyle{unsrt}

\end{document}